\documentclass{article}
\usepackage{cite}
\usepackage{amsmath,amssymb,amsfonts}
\usepackage{algorithmic}
\usepackage{graphicx}
\usepackage{textcomp}

\usepackage[utf8]{inputenc}

\begin{document}

\title{A dataset for complex activity recognition with micro and macro activities in a cooking scenario }

\author{Paula Lago$^1$,  Shingo Takeda$^1$,  Sayeda Shamma Alia$^1$, Kohei Adachi$^1$,\\ Brahim Bennai$^1$,Fran\c cois Charpillet$^3$, Sozo Inoue$^1$}

\date{$^1$ Department of Basic Science,Kyushu Institute of Technology,\\$^3$ INRIA, Nancy, France}

\maketitle

\noindent\textbf{Abstract}
Complex activity recognition can benefit from understanding the steps that compose them. Current datasets, however, are annotated with one label only, hindering research in this direction. In this paper, we describe a new dataset for sensor-based activity recognition featuring macro and micro activities in a cooking scenario. Three sensing systems measured simultaneously, namely a motion capture system, tracking 25 points on the body; two smartphone accelerometers, one on the hip and the other one on the forearm; and two smartwatches one on each wrist. The dataset is labeled for both the recipes (macro activities) and the steps (micro activities). We summarize the results of a baseline classification using traditional activity recognition pipelines. The dataset is designed to be easily used to test and develop activity recognition approaches.


\section{Introduction}
Activity recognition is the "automatic recognition of physical activities” from sensor data~\cite{Bulling2014}. The most common application of wearable sensor-based activity recognition is indeed in fitness trackers, but research has also focused on more complex activities such as daily living activities~\cite{Riboni2019709,8945220} or work-related activities~\cite{Inoue2016,DiPietro2019}. However, the accuracy of recognition for such complex activities is still low. This is due to several factors like the inter-class similarity, the difficulty in defining each activity and its boundaries, and the lack of open datasets.  

A complex activity is composed of actions~\cite{Liu2015} while physical activities are made of a periodic repetition of the same action. For example, walking entails the repetition of steps. An activity such as “cooking”, however, involves actions like “cut”, “take” or “mix” in different order and frequency. We call these actions “micro-activities” and we name the complex activities “macro activities”. Identifying micro-activities is useful for reasoning about macro-activities: to identify if all the required micro-activities were followed, to identify if they were in the correct order and to identify differences among macro activities or in the execution of any macro-activity. 

To show how the recognition of micro activities can aid reasoning, consider a nursing scenario. Nurses usually take blood from their patients. The steps involved are ``hand washing'' , ``opening injection'', ``blood collection'', ``healing'' and ``disposing waste''. In this example, nurses must wash their hands prior to any other step to prevent infections but they can change the order of the last two steps without much problem. For the activity ``changing diaper'', ``hand-washing'' must be done both at the beginning and at the end. Identifying micro-activities, as well as the macro-activity of which they  are part of, is an important step in reasoning about complex activities. 

Few open datasets have more than one level of labels~(Section~\ref{sec:related}) despite the importance of these relations. In this paper, we present a new dataset with two-level labels: one level for the macro activities and the other for the micro-activities~(Section~\ref{sec:summary}). The main contributions in this dataset are the multi-modality and multi-sensor data and the two-level labeling approach. The data was collected in a cooking scenario using recipes as macro-activities. The micro activities, steps in a recipe, appear in a different order and frequency depending on the macro activity.  We chose this setting because cooking is an important activity of daily living. The recognition and analysis of cooking activities can offer insights on health, well-being and the ability to live independently. In addition, it offers an easy setup to study the relation between micro and macro activities.

We analyze the use of traditional activity recognition pipelines~\cite{Bulling2014} for the recognition of micro-activities (Section~\ref{sec:baseline}). We show that the recognition of micro-activities must consider new models different from those of physical activity recognition because of the lack of periodicity, data imbalances and differences among subjects.

\section{Related datasets}
\label{sec:related}

Although recently there has been increasing interest for sharing datasets, the diversity and compatibility of open datasets remains a problem. Most datasets with wearable devices focus on locomotion activities. From the datasets focusing on other activities, the diversity of activities is high, making it difficult to combine different datasets into a single one with the same activities.  A review of sensor-based activity recognition datasets can be found in~\cite{De-La-Hoz-Franco2018}.

As no one dataset can cover all possible scenarios, or have enough participants, we believe that creating compatible dataset, with similar sensors and similar activities is a good way of creating collaborative data. With similar sensors and activities, it will be possible to combine several datasets for different research purposes. 

Our dataset thus aims to create an additional source for researchers looking into cooking activity recognition, gesture recognition, and other scenarios that might benefit from the different granularity of activity labels.
We use several sensors including optical motion capture to make it diverse. Although it has a small number of participants and recipes, the large number of sensors gives the dataset a special recognition since other datasets lack body positions or other data that can help to make sense of the data. 

In this section, we describe other datasets that are publicly available and comprise cooking scenarios and/or micro and macro activities.

\textbf{CMU Lifestyle dataset:}~\cite{cmudataset} This dataset was recorded using three different modalities including: video, audio, and 5 inertial measurement units placed on the subjects back, legs, and arms. The main dataset contains 18 subjects cooking five different recipes: brownies,pizza, sandwich, salad and scrambled eggs. Labels are given for detailed activities including their object, for example, ``put-bakingpan-into-oven'', ``walk--to-counter'' or ``open-browniebag''. Although the dataset planned to use motion capture, it is only available for one subject. 

\textbf{Cooking activities dataset}~\cite{cookingdataset}: The data was collected using motion capture system based on wearable inertial measurement units (five positions). The scope was the activities done during meal time, labeling different micro activities and five macro activities: cooking meal, setting table, eating meal, cleaning up and putting away utensils. The subjects of this experiment followed an experimenter who described the tasks, so there was a slight dependency although subjects were free to chose their order of actions. Only one recipe is done. 

\textbf{Opportunity dataset}~\cite{5573462}: This benchmark dataset contains information about gestures that occur during some high level activities, similar to micro and macro activities. The data was collected in an environment rich with sensors. There were 72 sensors in the environment and on the body of the subjects. The dataset includes complex cooking activity mainly associated with breakfast (preparing a sandwich). Four subjects performed 17 micro activities in a predefined scenario. The main limitation is the number of recipes (only one). 

\textbf{Unmodified Kitchen Dataset}~\cite{Mohammad2017} This dataset was collected in real kitchens with 10 women participants. It consists of 2 recipes cooked freely, and 74 basic activity labels given for each hand separately. Accelerometer data from two smartwatches, one in each wrist, was collected. The labeled activities include ``Mix with chopsticks'', ``Move aside'' and ``Shake''. This dataset provides realistic data both in terms of the preparation of recipes and data, which was collected using commercial devices. The limitation is in the number of recipes and, due to the detailed granularity of labels, small number of samples for most of them.

\section{Data Collection Experiment Design}
\label{sec:experiment}
\begin{figure}[hbt]
     \centering
    \includegraphics[width=0.75\linewidth]{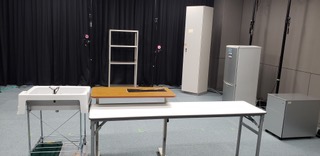}
    \includegraphics[width=0.75\linewidth]{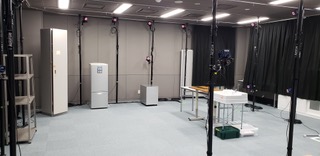}
     \caption{Smart Life Care Unit of Kyushu Institute of Technology equipped as a kitchen. View from the front (top) and side (bottom)}
     \label{fig:studio}
\end{figure}

The dataset was recorded in the Smart Life Care Unit of the Kyushu Institute of Technology in Japan~\cite{openlaburl}. This unit is located in the Wakamatsu Campus of the University and has optical and inertial-sensor based motion capture equipments, sensors such as EMG, EEG, eye movement sensors and others and was equipped with furniture so as to simulate a kitchen environment as shown in Figure~\ref{fig:studio}. The experiment was held on November 19, 22,and 25, 2019.

In this section, we detail the activities, sensors, data collection protocol used in the experiment. 

\subsection{Activities}
We recorded data for 10 micro activities and 3 macro activities. One macro activities corresponds to one recipe. For the success of the experiment, we had the following restrictions when choosing the recipes:
\begin{itemize}
    \item Recipes should not involve heating as it can not be done in the laboratory for safety. 
    \item Each micro activity should be present in at least 2 macro activities. 
    \item The duration of the recipe should be short and it should be easy to prepare as volunteers have no cooking experience. 
\end{itemize}

Considering the previous restrictions, we chose the following recipes:
\begin{itemize}
    \item \textbf{\textit{Sandwich}} A cheese and vegetables (lettuce and tomato) sandwich. Although it includes a 'spread' micro activity for spreading mayonnaise, we don't consider it as it is not a part of any other activity. 
    \item \textbf{\textit{Cereal}}: Pouring milk and cereal into a bowl. We include banana into the cereal to include cut and peel micro activities. 
    \item \textbf{\textit{Fruit salad}}: A fruit salad including 3  fruits (banana, apple and tangerine) that must be peeled, cut and then mixed with yogurt. 
\end{itemize}

\begin{figure}
    \centering
     \includegraphics[width=0.95\linewidth]{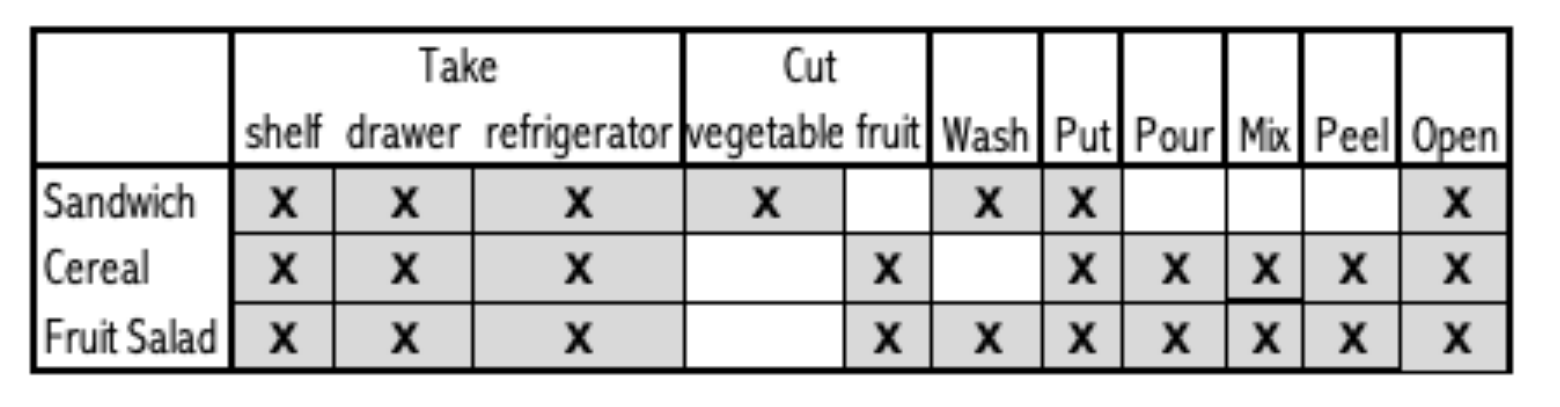}
     \caption{Micro activities involved in each macro activity}
     \label{fig:matrix}
 \end{figure}

We designed the micro-activities in accordance to previous datasets labels~(Section~\ref{sec:related}). In those datasets, the action is accompanied by the object, if it is relevant. To make our dataset compatible with previous ones, we also collect object information as part of the micro activity label. Figure~\ref{fig:matrix} shows a summary of the micro activities involved in each recipe. 

As a micro activities has a semantic meaning instead of a motion related meaning, we expect some of them to show some intra-class variability in the sensor data. For instance, the peel activity is expected to have very different movements depending on the object of the activity, i.e. peeling a banana is significantly different than peeling an apple. However, it is the semantic meaning that is interesting to recognize. Similarly, cutting can be slightly different when cutting fruits than when cutting vegetables due to different forces applied. As another example, we expect take to differ significantly across the three objects (shelf, drawer and refrigerator). Please note that in this dataset ``Take'' involves opening, taking and closing the container, whereas in other datasets such as the CMU dataset, this would be labeled as three actions. In our case, ``Open'' refers to opening a food package such as milk carton or yogurt pot. 

Other micro-activities were labeled but are not considered as they appear in only one of the activities. This include ``Open'', ``Spread'' and ``Add''.

The following is a description of each activity.
\begin{itemize}
    \item \textit{Peel}: Removing the skin of a fruit. In this dataset it can refer to peeling a banana, apple or tangerine. For the apple, a knife is used whereas for the banana and tangerine the hands are used.  
    \item \textit{Take}: Taking an ingredient or object from either the shelf, refrigerator or drawer. For the shelf, raising the arms might be involved depending on the height of both the object's location and the participant. For taking from the drawer, a way of going low (either bending or squatting) is required. 
    \item \textit{Wash}: A simulation of washing a fruit or vegetable. Since there was no real water connection for the sink, this activity was mainly simulated by opening the water, placing the object beneath and closing. 
    \item \textit{Cut}: Cutting a fruit (banana, apple, tangerine) or vegetable (lettuce, tomato) 
    \item \textit{Pour}: Pouring yogurt from pot to bowl or pouring milk from carton to bowl.  
    \item \textit{Put}: Including one ingredient into a mix of the others. This includes, putting cheese on top of bread, putting tomato or lettuce for the sandwich; putting fruits into the bowl for fruit salad; and, putting cereal into the bowl. 
    \item \textit{Mix}: This means combining fruits in the salad with a spoon or cereal with the milk. 
    \item \textit{Open}: Includes opening packages like milk carton, cheese slice (packaged individually), and yogurt pot. 
    \item \textit{other}: It refers to micro-activities that were not part of at least two recipes and were not considered as labels for the final dataset. For example, 'Spread'. It might also refer to the static poses at the beginning and end of the recordings. 
\end{itemize}

\subsection{Sensors}
We collected data from motion capture, open pose and accelerometer sensors. Video was also be recorded for each run, but it is not released due to privacy concerns of the participants. The description of each data source is given below.

\begin{figure}
    \centering
    \includegraphics[width=0.9\textwidth]{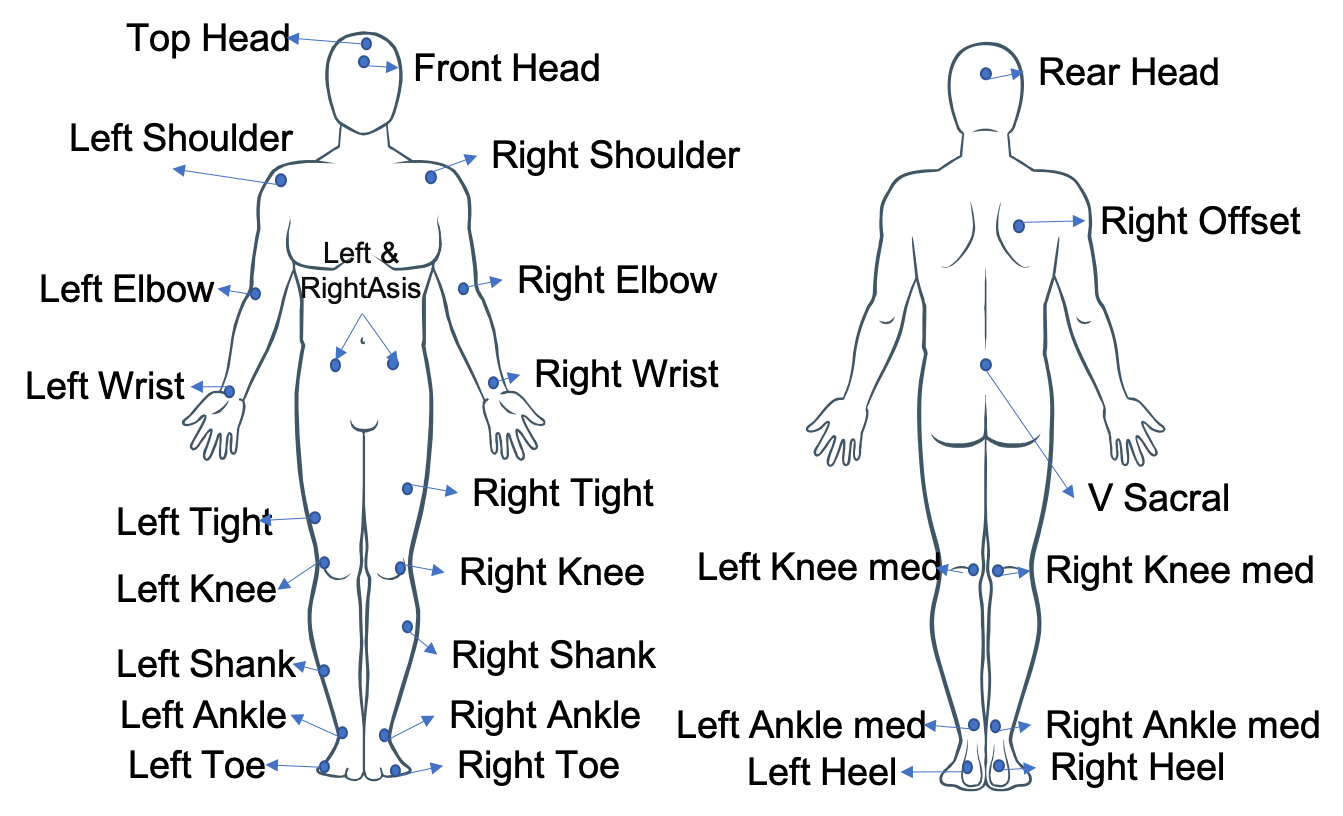}
    \caption{Motion capture markers used in this dataset}
    \label{fig:markers}
\end{figure}

\textbf{Motion Capture:} We used the motion capture system form Motion Analysis Company~\footnote{\underline{http://motionanalysis.com/movement-analysis/}}. The setup used consisted of 29 body markers located as in Figure~\ref{fig:markers}. The markers are tracked using 16 infrared cameras (Kestrel Digital Real Time). The three dimensional position of each marker was recorded with a frequency of 100Hz. The markers may be labeled incorrectly in some cases due to the complex setting which sometimes obstructs some markers. 

\textbf{Open Pose:}~\cite{cao2018openpose} Open Pose is an open source system to detect 135 keypoints of the human body, hand, facial, and foot using a single COTS camera. We use the keypoints from body (25). The main purpose is to compare results of motion capture with those of open pose. Open-pose may be used as a low-cost, low-precision body-tracking system.  

\begin{figure}
    \centering
    \includegraphics[width=0.7\linewidth]{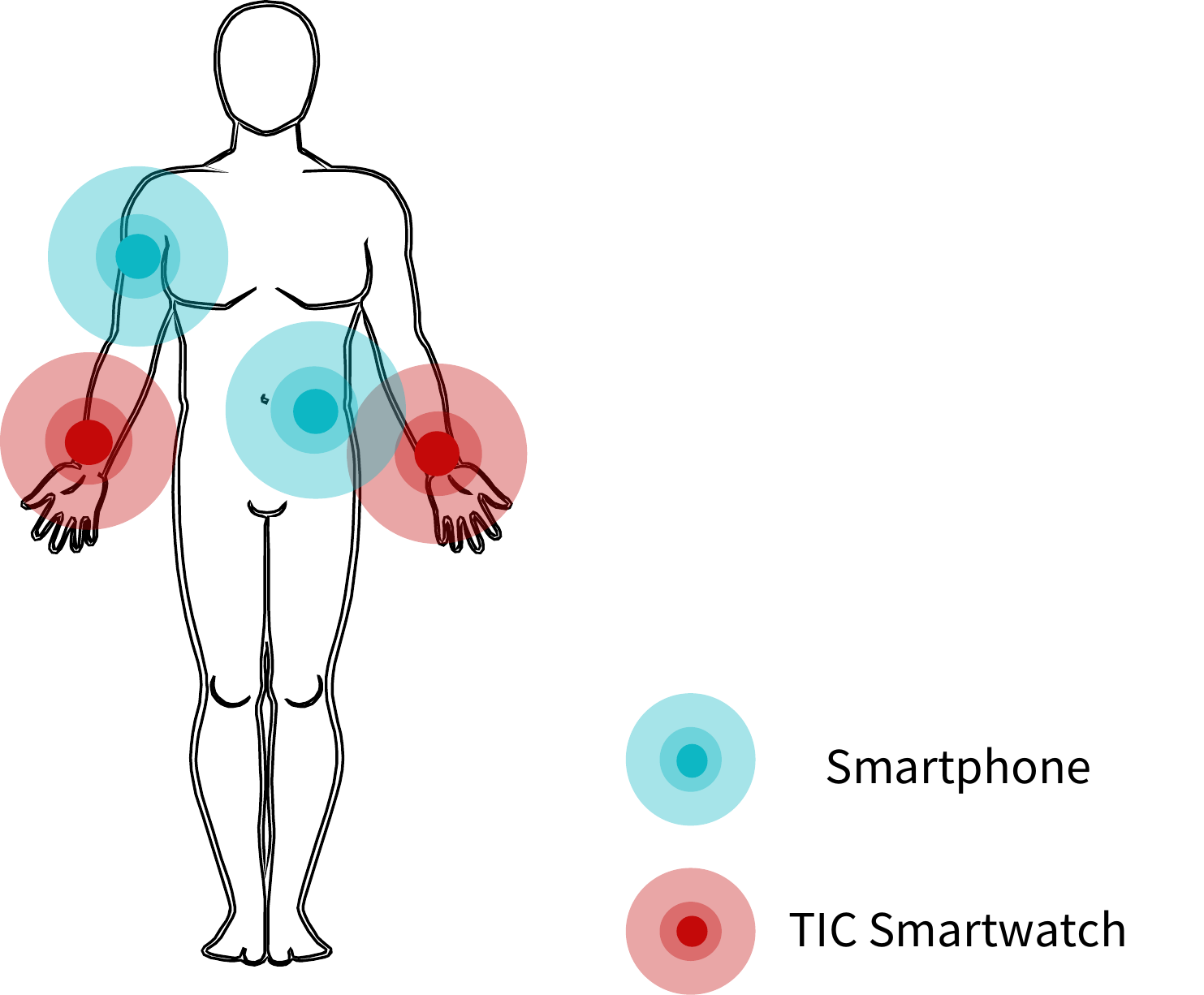}
    \caption{Placement of the acceleration sensors for the experiment}
    \label{fig:accelerometers}
\end{figure}

\textbf{Accelerometer sensor:} As accelerometer sensors, we used two smartphones and two smartwatches placed as in Figure~\ref{fig:accelerometers}.
The smartphones have a sensitivity of $\pm1g$ and a sampling rate of ~70Hz in average. The smartwatches are a TIC Watch E, which uses Google wear and connects to the smartphone via bluetooth. The sampling rate was set to the maximum possible, but it varied greatly during the experiment from ~50Hz to ~250Hz. Their sensitivity is of $\pm2g$.  All measurements are given in $m/s^2$

Two smartphones were used. The left hip smartphone (connected to the left wrist smartwatch) was a Samsung Galaxy S9 SCV38 and the right arm smartphone (connected to the right wrist smartwatch) was a Huawei P20 Lite smartphone. Both smartphones used Android version 9 as operating system.


\subsection{Protocol}
During the experiment, there was a director, four observers and one subject. The director read the next steps of the recipe for the subject to perform the activity in the desired order and without forgetting any step. Each observer controlled one sensor (smartphone X2, motion capture and video) to start and stop the sensing at the same time. Only one observer labeled each step with an application designed for this purpose. 
\begin{figure}
    \centering
    \includegraphics[width=0.9\linewidth]{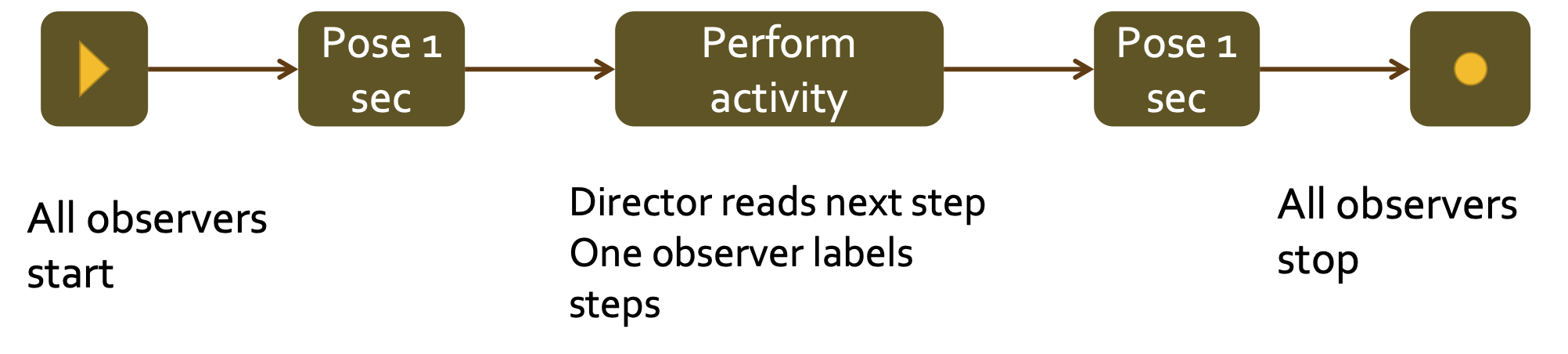}
    \caption{Protocol to record each trial in the experiment}
    \label{fig:protocol}
\end{figure}

Based on previous experiences in data collection experiments, the following practices were followed. 

\textit{Sensor synchronization} was made by clock synchronization. We also use initial and end pose calibrations so that the start of the action can be better synchronized. 

\textit{Orientation calibration pose} In many datasets, the orientation of the sensor is not known. This makes the analysis difficult since the measurements depend on the orientation. We use an initial pose holding for 0.5 seconds so that the initial data can be used for orientation calibration of all accelerometers. 

\subsection{Participants}
All four participants are male students between 18 and 25 years who participated voluntarily and without compensation in this experiment. 


\section{Data Summary}
\label{sec:summary}

In this section we describe statistics of the dataset. We first describe statistics related to the time distributions of the dataset. 
Then we analyze the accelerometer sensor reading distributions to analyze the differences and similarities in the distribution for each micro-activity. 
Finally, we analyze the quality of the data by estimating the label quality, the  missing data rate, and the timestamp synchronization quality. 

\subsection{Analysis by time}
\label{sec:time}
The following analysis aims to answer the following questions:\newline
\textit{\bf TQ.1: How much time was recorded for each activity?}\newline
\textit{\bf TQ.2:  How much time was recorded for each subject?}\newline
\textit{\bf TQ.3: What are the average durations of each activity?}\newline

\begin{figure}
    \centering
    \includegraphics[width=0.9\linewidth]{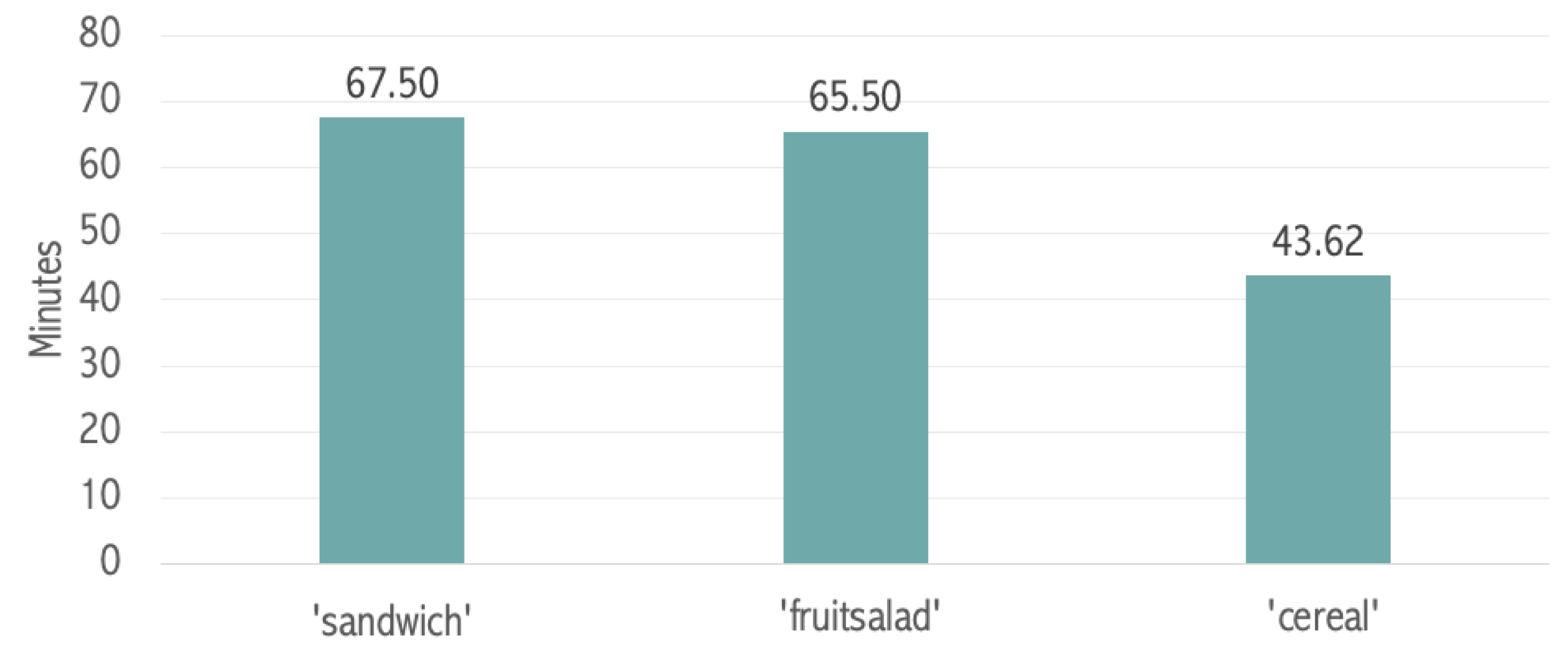}
    \caption{Time distribution by recipe}
    \label{fig:time_recipe}
\end{figure}

\begin{figure}
    \centering
    \includegraphics[width=0.9\linewidth]{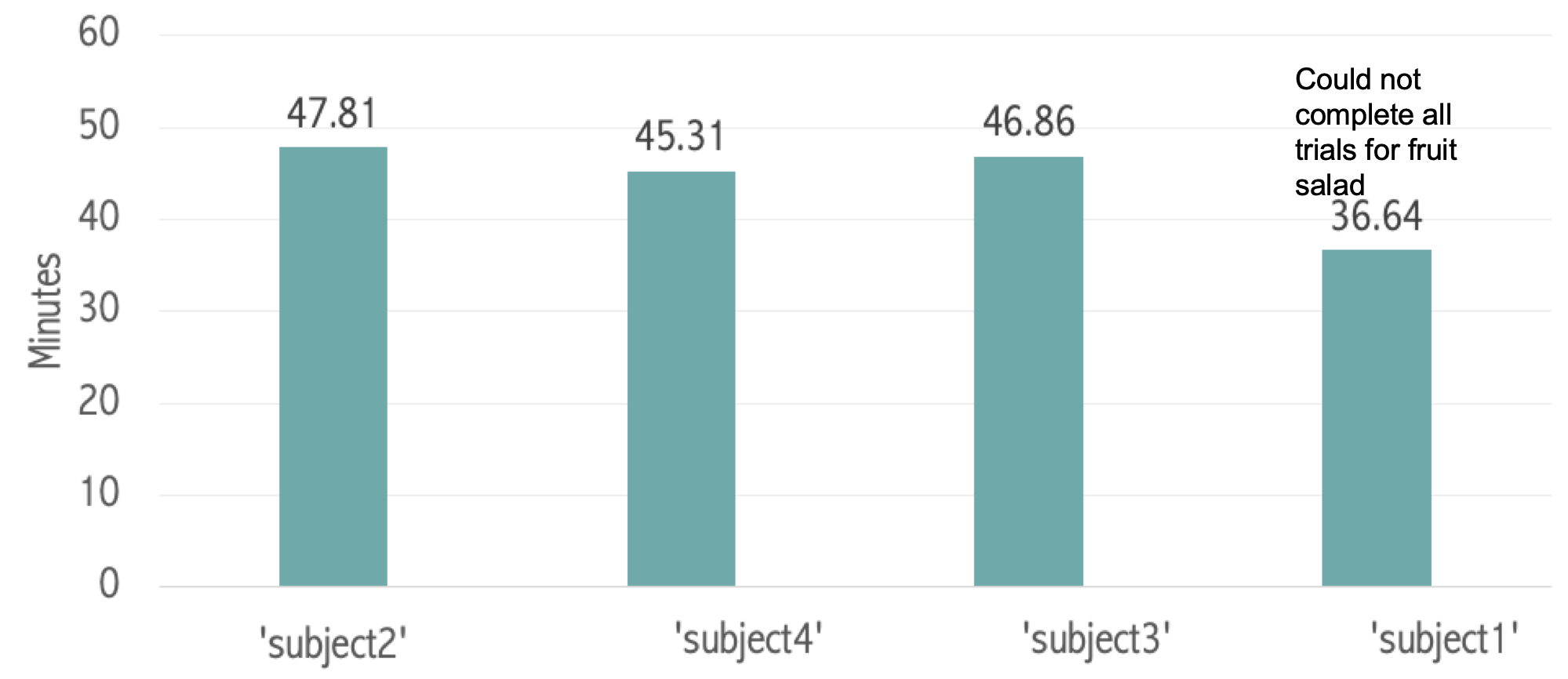}
    \caption{Time distribution by subject}
    \label{fig:time_subject}
\end{figure}

The dataset comprises almost three hours of data, almost evenly distributed across each recipe~(Figure~\ref{fig:time_recipe}) and across each subject~(Figure~\ref{fig:time_subject}). Only one subject (subject 1) has less recorded time due to an error during the experiment. For this reason, Subject 1 has only 2 trials for the fruit salad activity. For all other recipes and all the other subjects, there are 5 trials per subject per recipe.  

\begin{figure}
    \centering
    \includegraphics[width=0.9\linewidth]{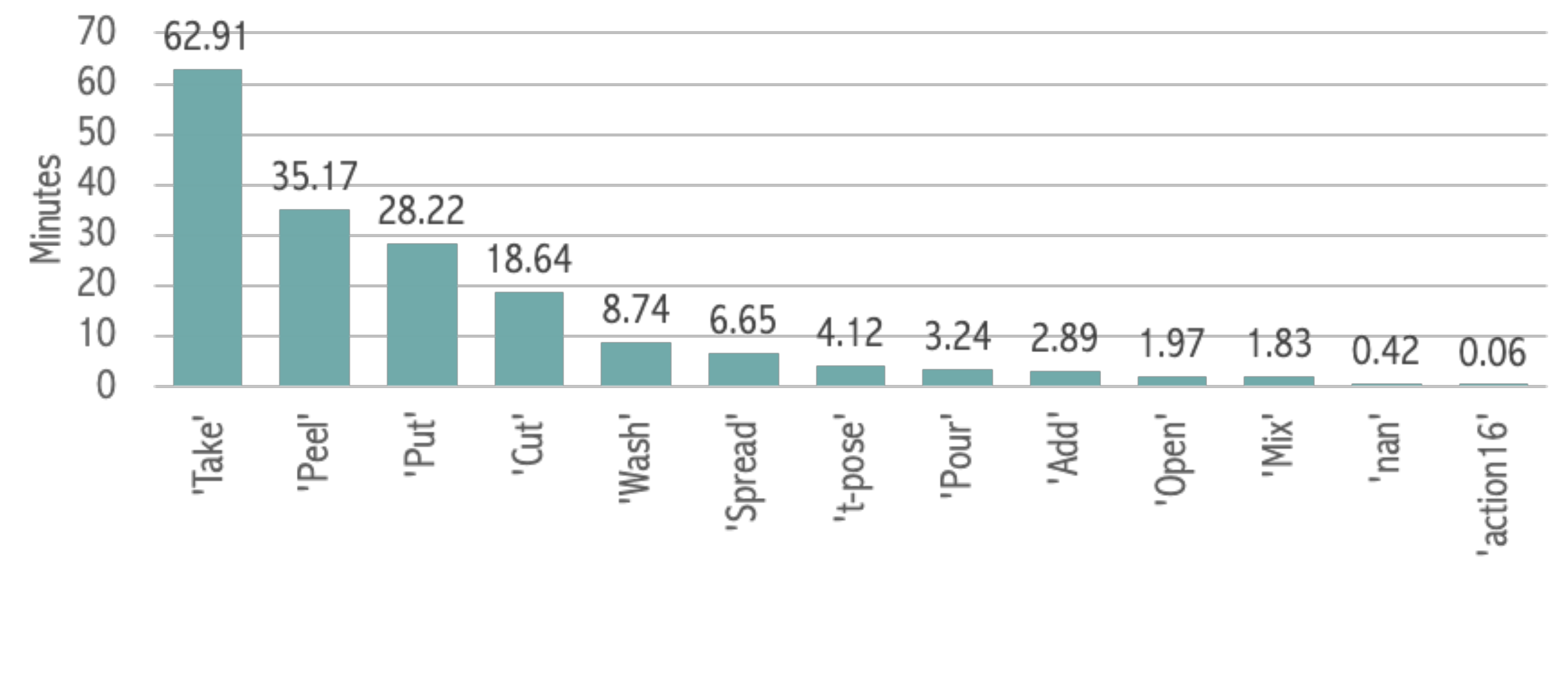}
    \caption{Time distribution by micro activity}
    \label{fig:time_microactivity}
\end{figure}

When looking at the time distribution across micro-activities~(Figure~\ref{fig:time_microactivity}), the dataset is imbalanced. The activities 'Take', 'Peel', 'Put', and 'Cut' make almost 2/3 of the data. The 'Take'  activity is the most frequent one and the 'Peel' activity was the most difficult to perform, so they take the longest time. 

The average duration of each activity is shown in~Table~\ref{tab:avg_duration}. It confirms that 'Peel' is the activity with the longest duration, due to its difficulty. The shortest activity is 'Open'. 

\begin{table}[]
\caption{Average duration of each micro activity}
    \label{tab:avg_duration}
    \centering
    \begin{tabular}{|c|c|c|}
    \hline
         \textbf{Activity} & \textbf{Avg. duration (seconds) } & \textbf{SD. of duration (seconds) }\\
         \hline
        ``Peel''&25.42&18.32\\
        ``Spread''&16.62&2.65\\
        ``Take''&13.98&3.48\\
        ``Wash''&10.92&3.54\\
        ``Cut''&8.74&3.48\\
        ``Pour''&8.56&2.26\\
        ``Put''&7.49&6.20\\
        ``Mix''&6.10&1.90\\
        ``Open''&4.93&0.79\\
        ``other''&6.70&5.95\\
        \hline
    \end{tabular}
\end{table}

\subsection{Accelerometer readings analysis}
\label{sec:measurements}
The following analysis is based only on the accelerometer sensor values and aims to answer the following questions:\newline
\textit{\bf MQ.1: Do the accelerometer readings have different distributions for each activity?}\newline
\textit{\bf MQ.2: What statistical measures will most likely aid in the classification process?}\newline

\begin{figure}
    \centering
    \includegraphics[width=0.3\textwidth]{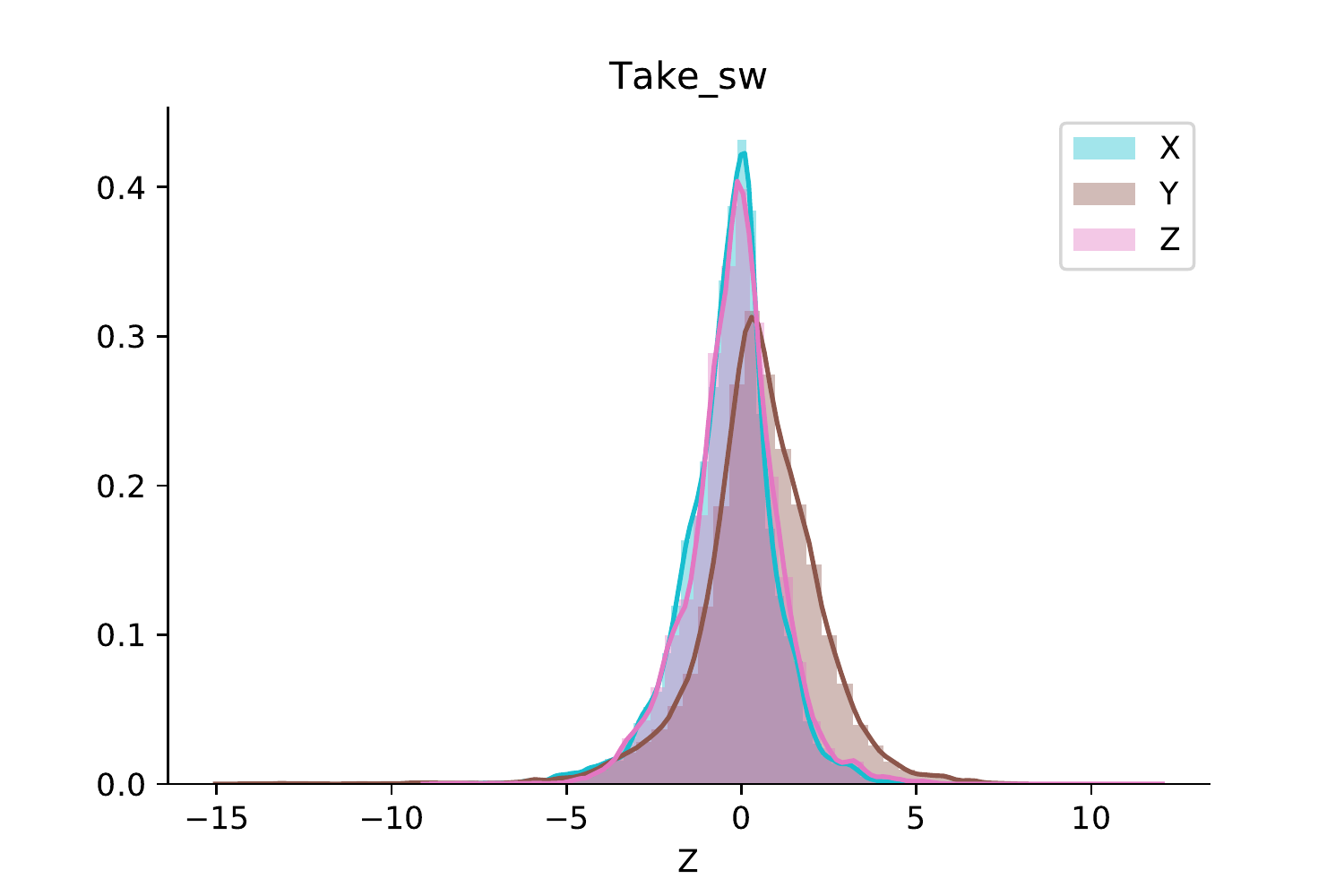}
    \includegraphics[width=0.3\textwidth]{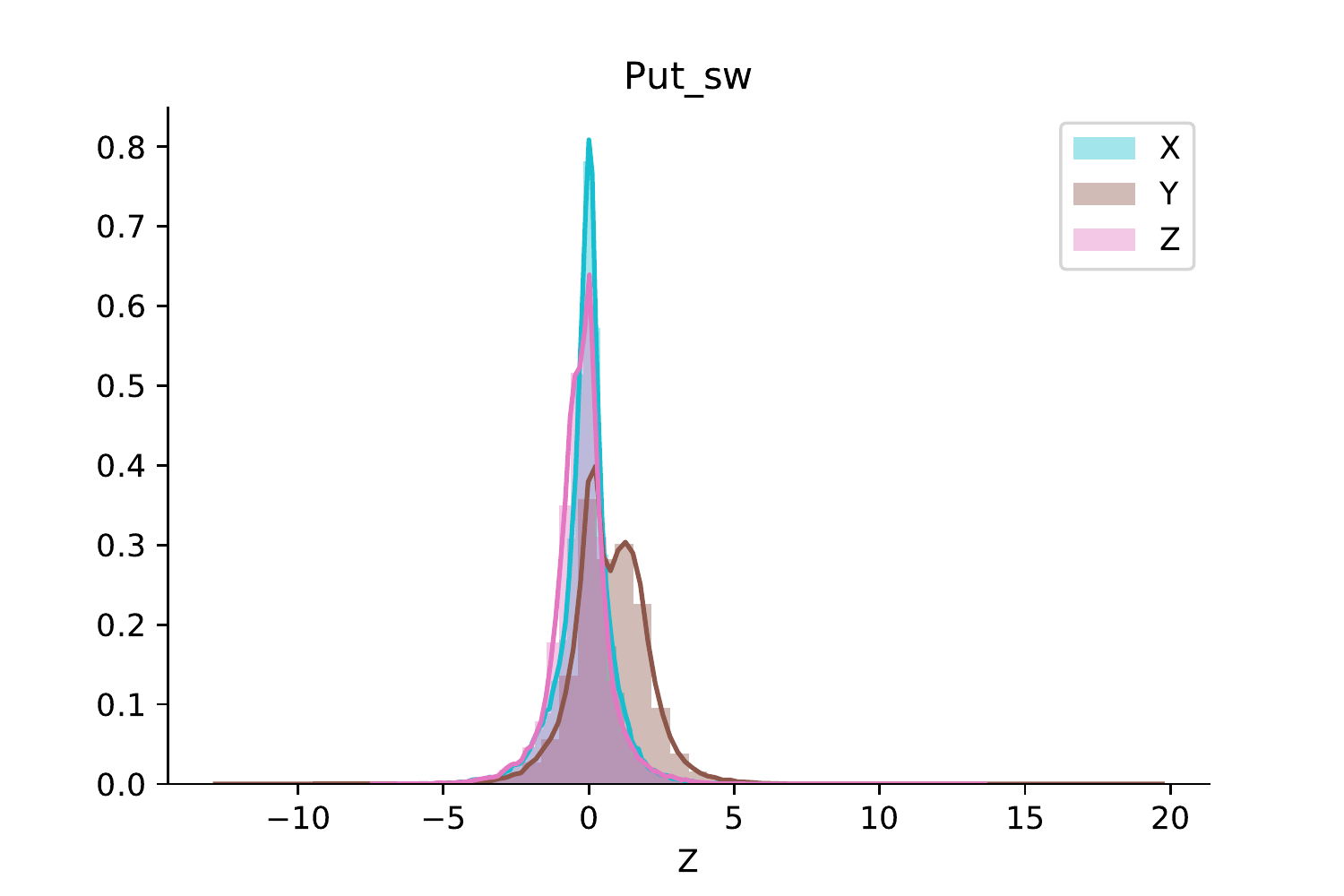}
    \includegraphics[width=0.3\textwidth]{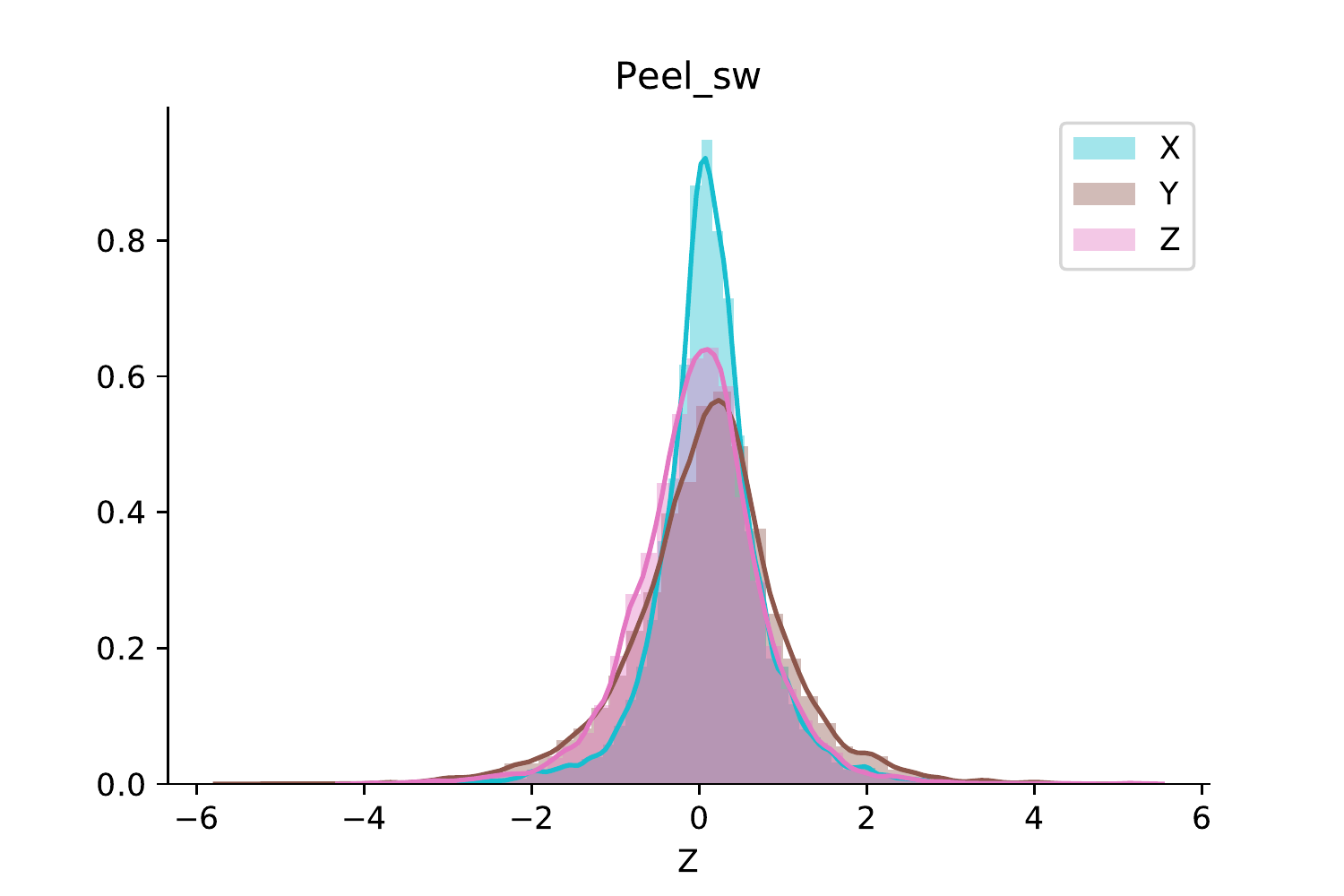}
    \includegraphics[width=0.3\textwidth]{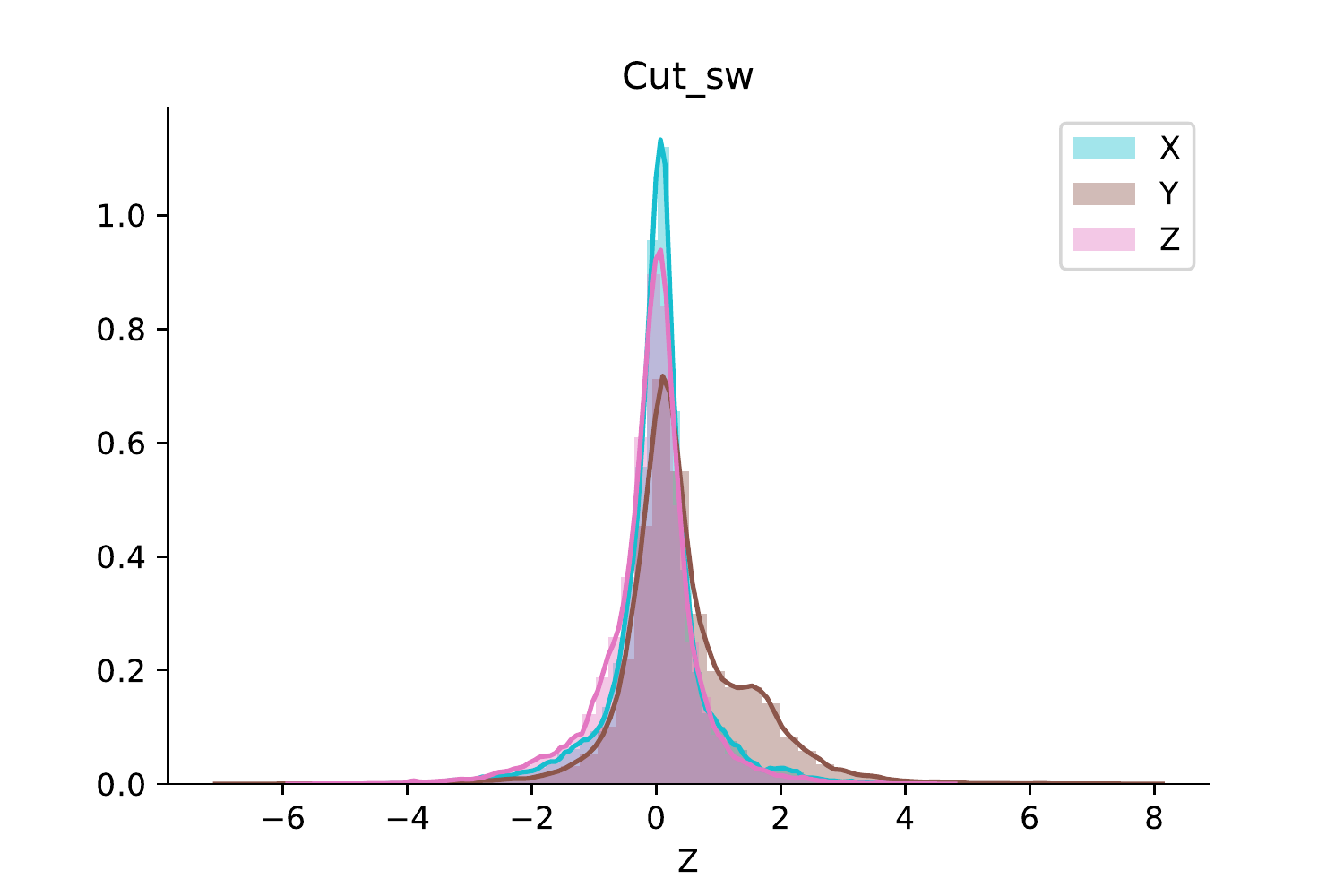}
    \includegraphics[width=0.3\textwidth]{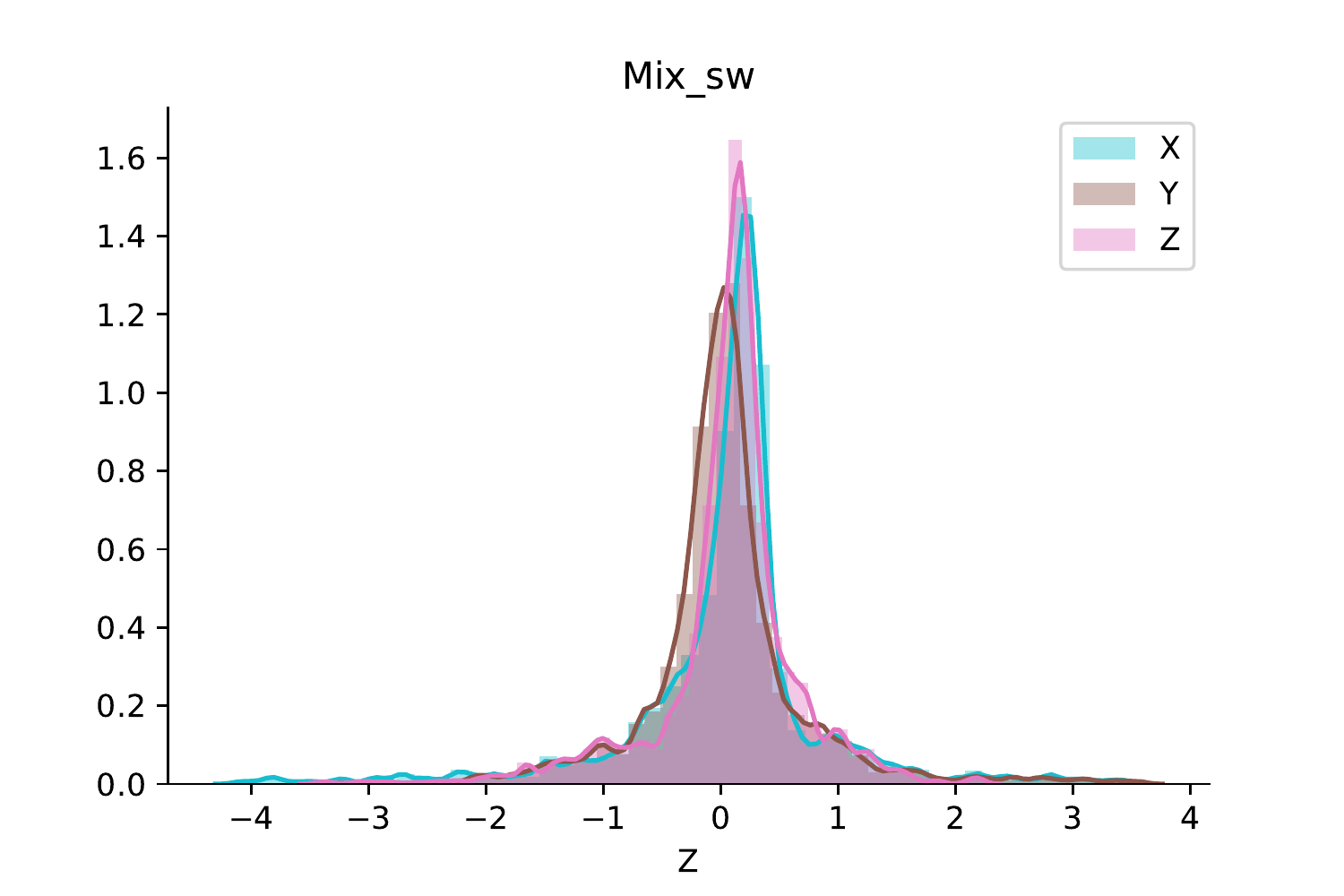}
    \includegraphics[width=0.3\textwidth]{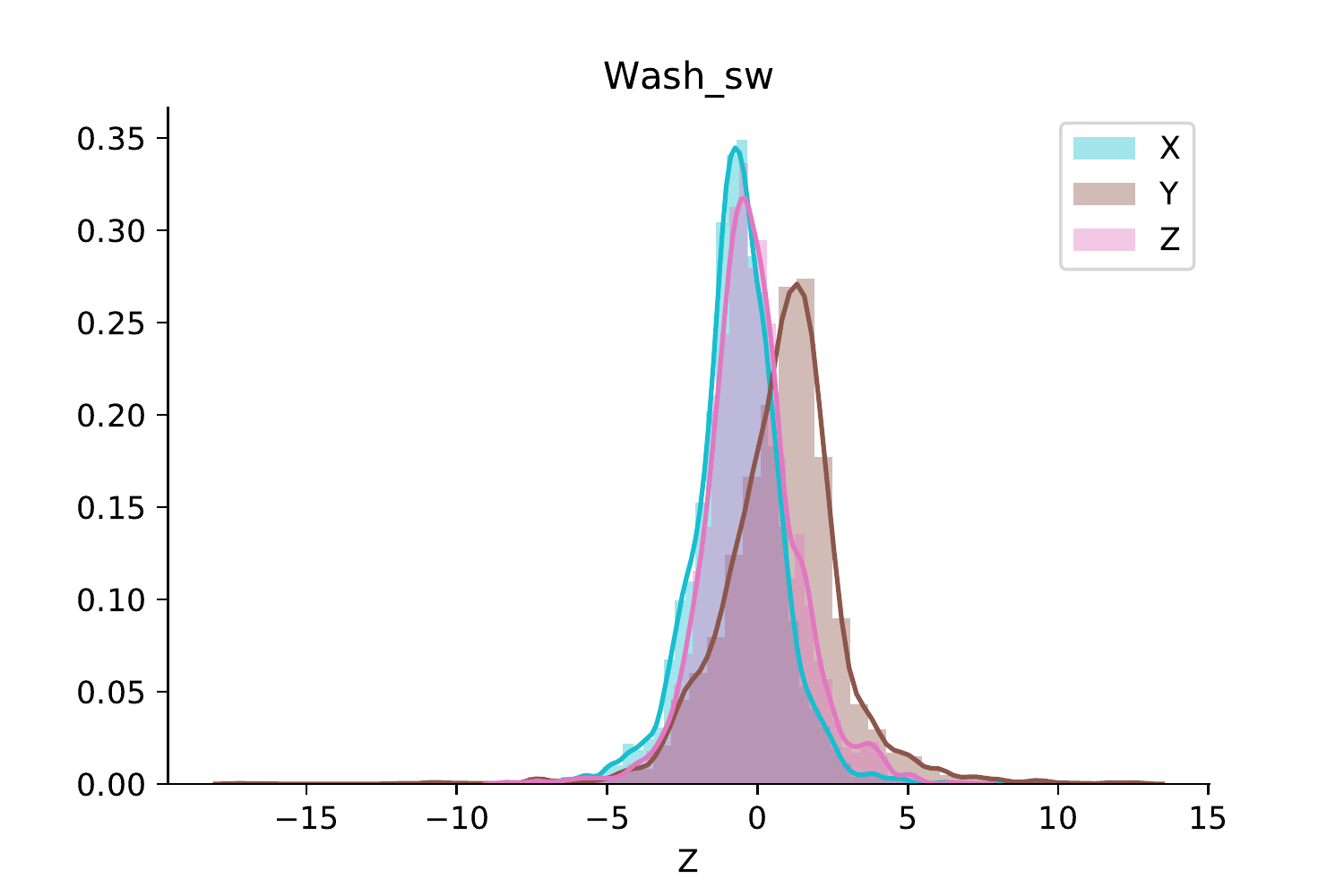}
    \caption{Distribution of the sensor measurements during different micro activities on the left wrist}
    \label{fig:dist_lw}
\end{figure}

\begin{figure}
    \centering
    \includegraphics[width=0.3\textwidth]{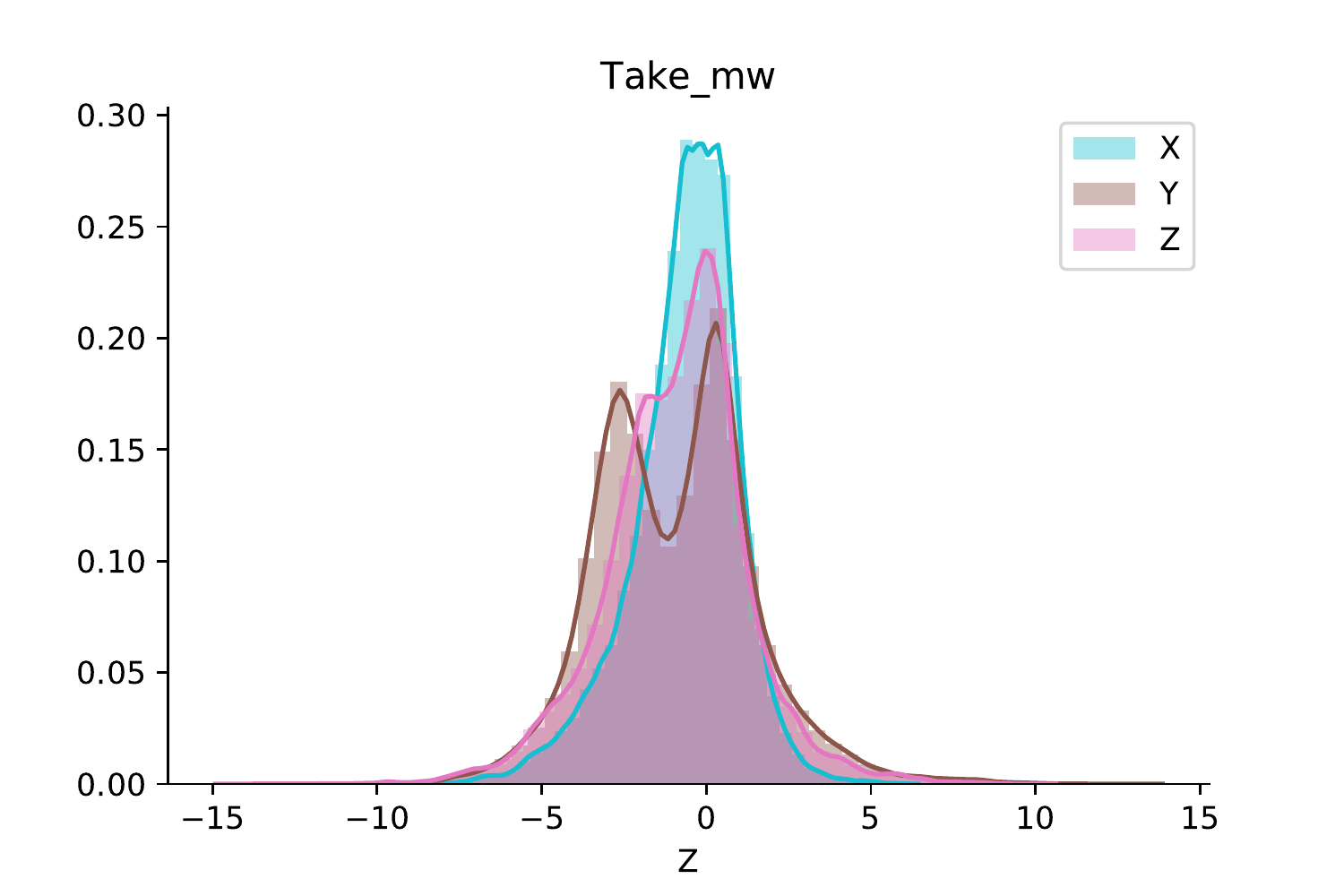}
    \includegraphics[width=0.3\textwidth]{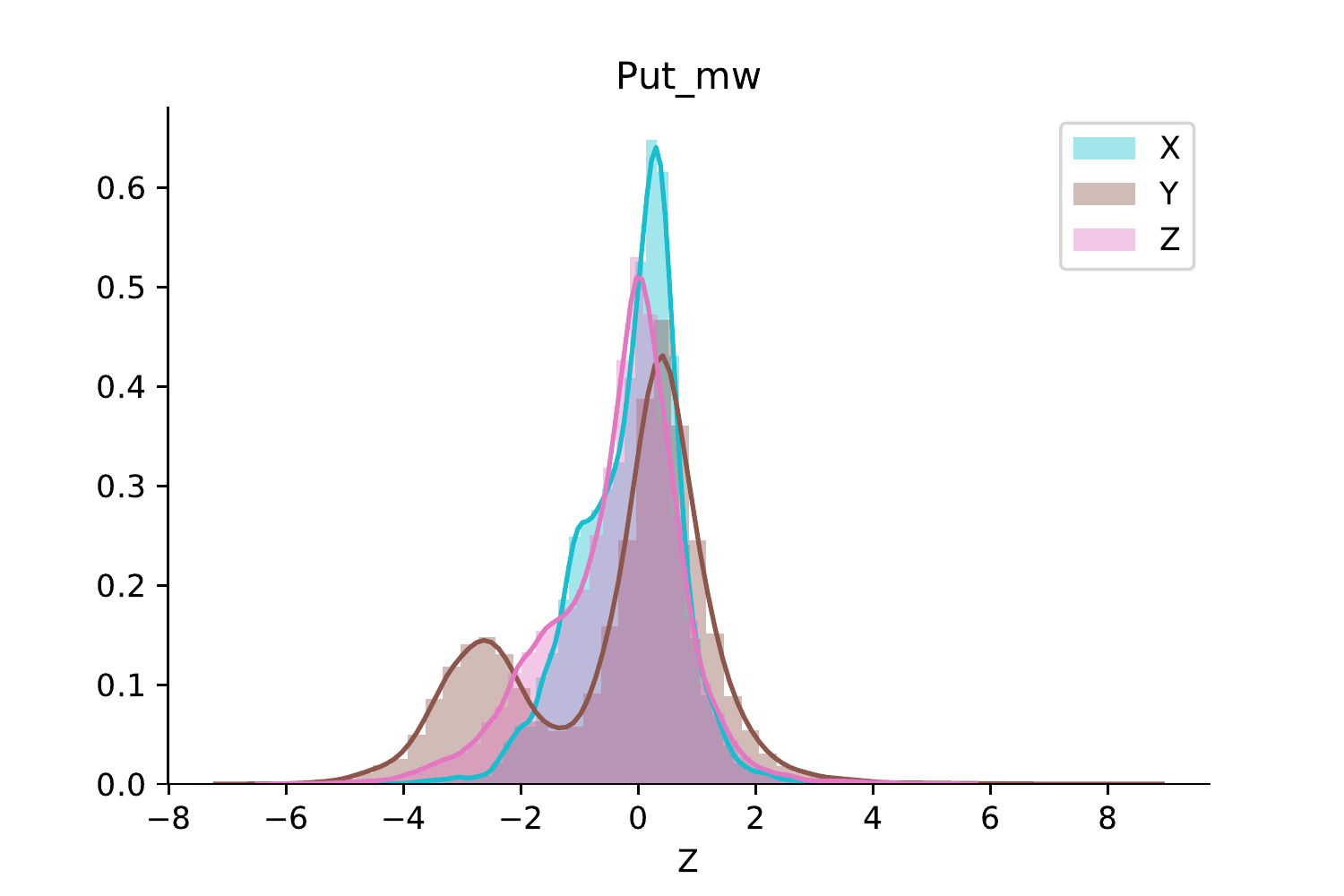}
    \includegraphics[width=0.3\textwidth]{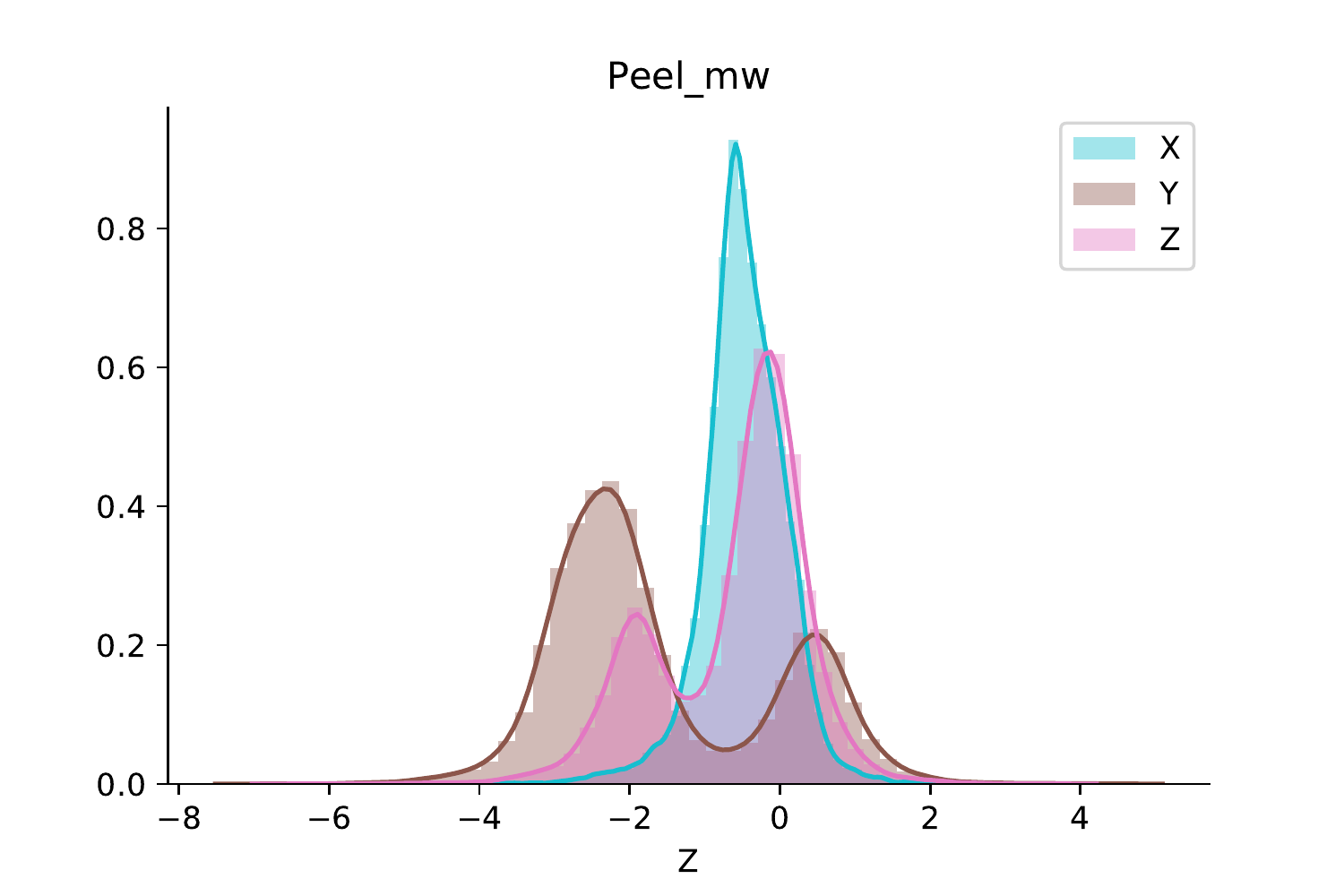}
    \includegraphics[width=0.3\textwidth]{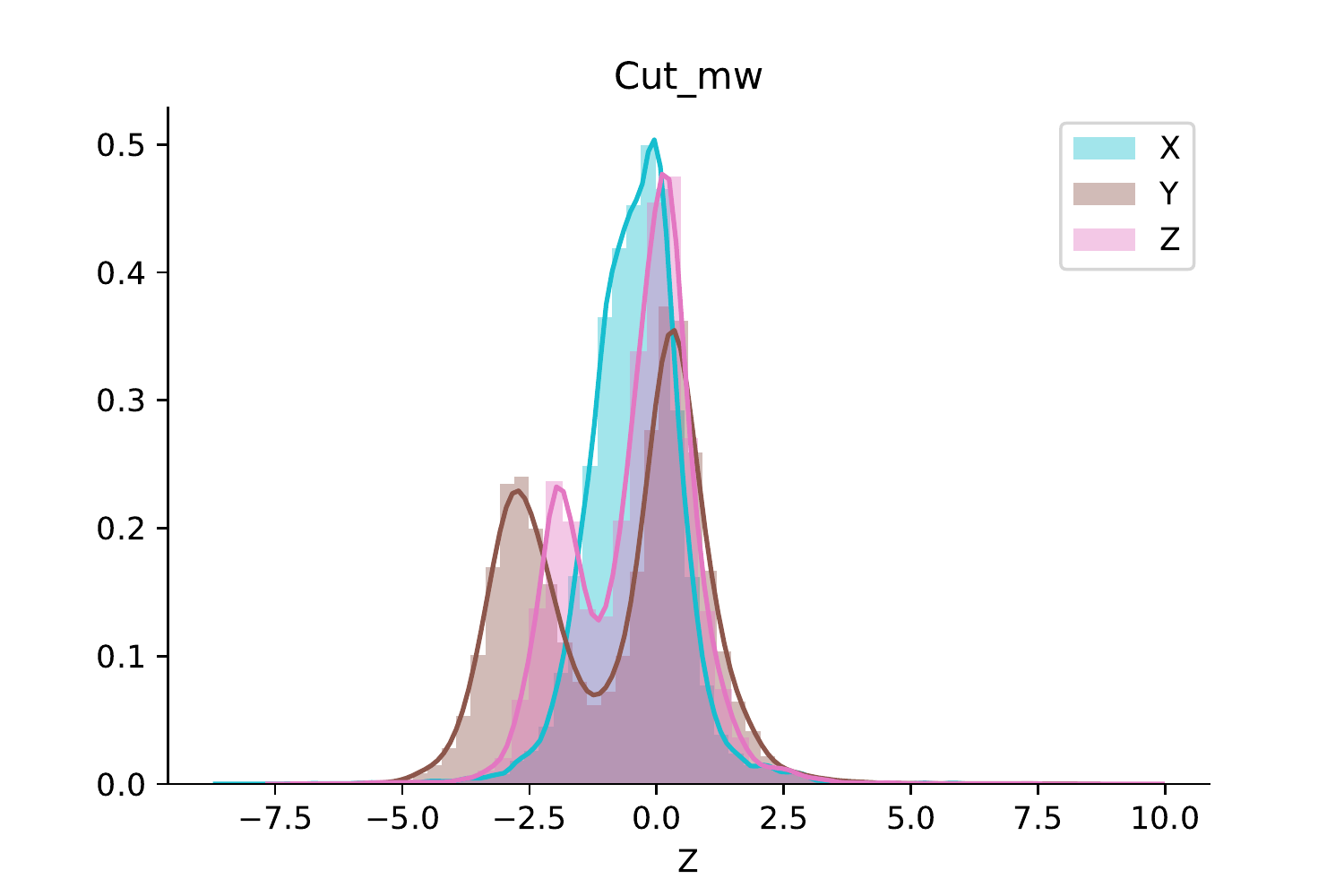}
    \includegraphics[width=0.3\textwidth]{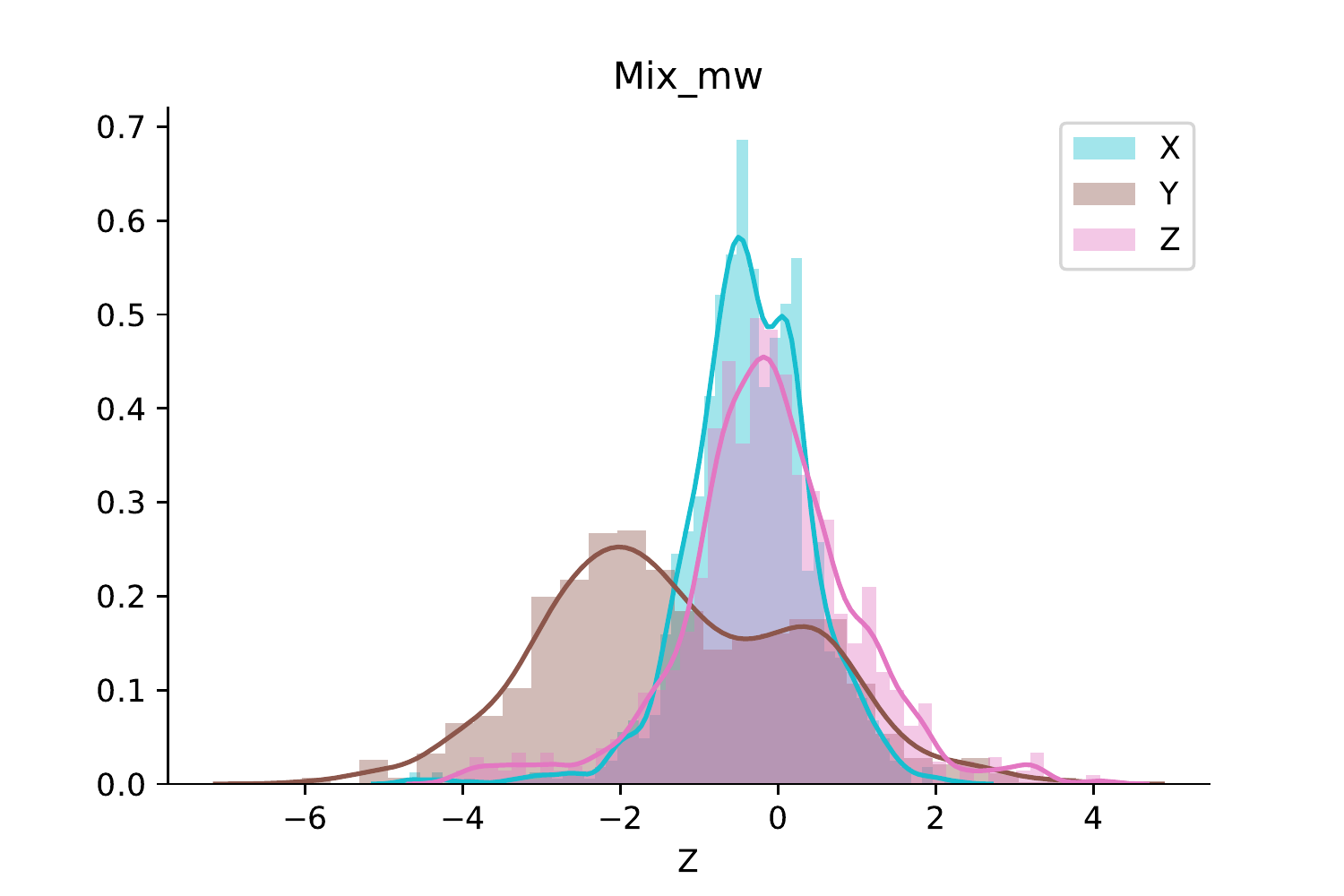}
    \includegraphics[width=0.3\textwidth]{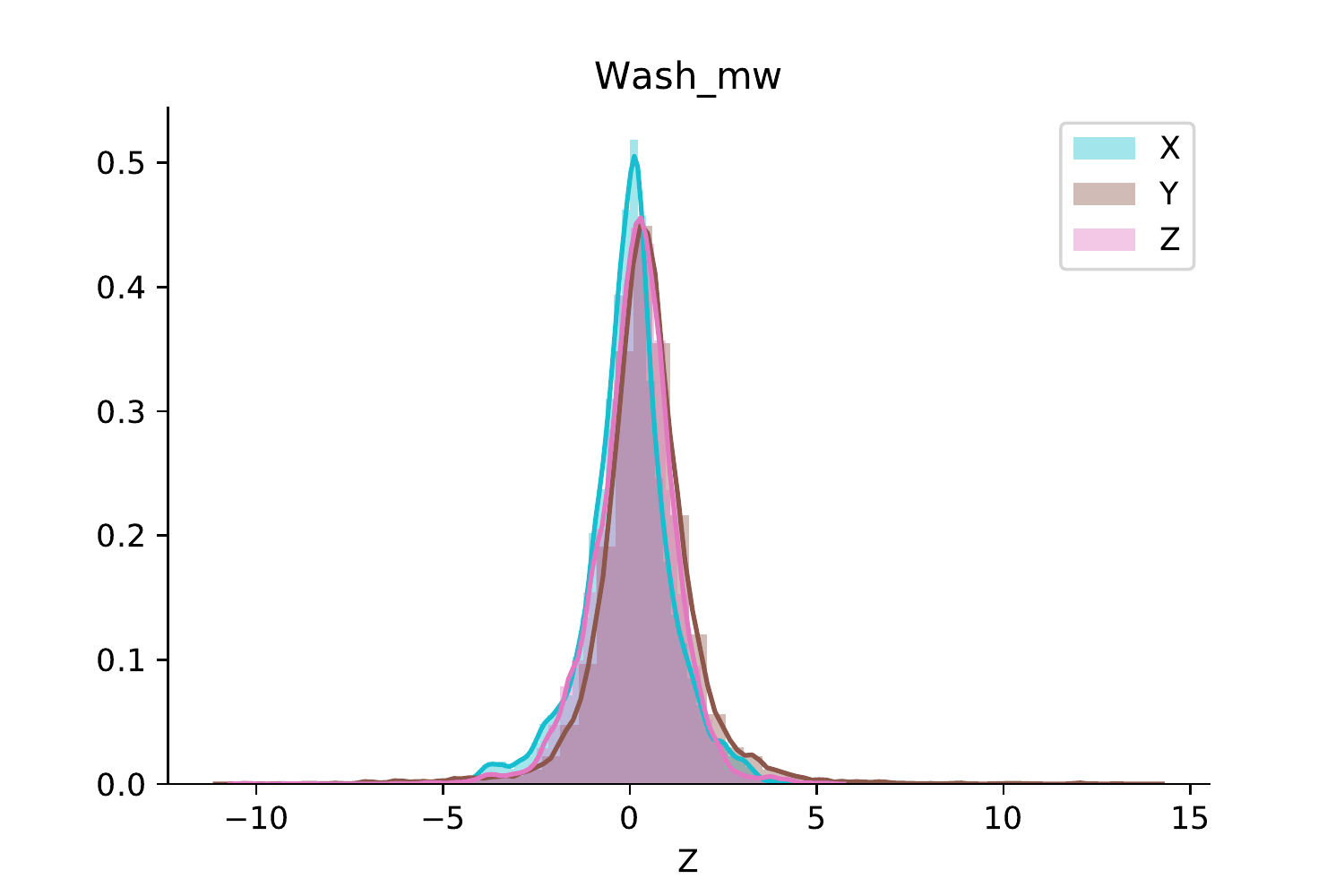}
    \caption{Distribution of the sensor measurements during different micro activities on the right wrist sensor}
    \label{fig:dist_rw}
\end{figure}

\begin{figure}
    \centering
    \includegraphics[width=0.3\textwidth]{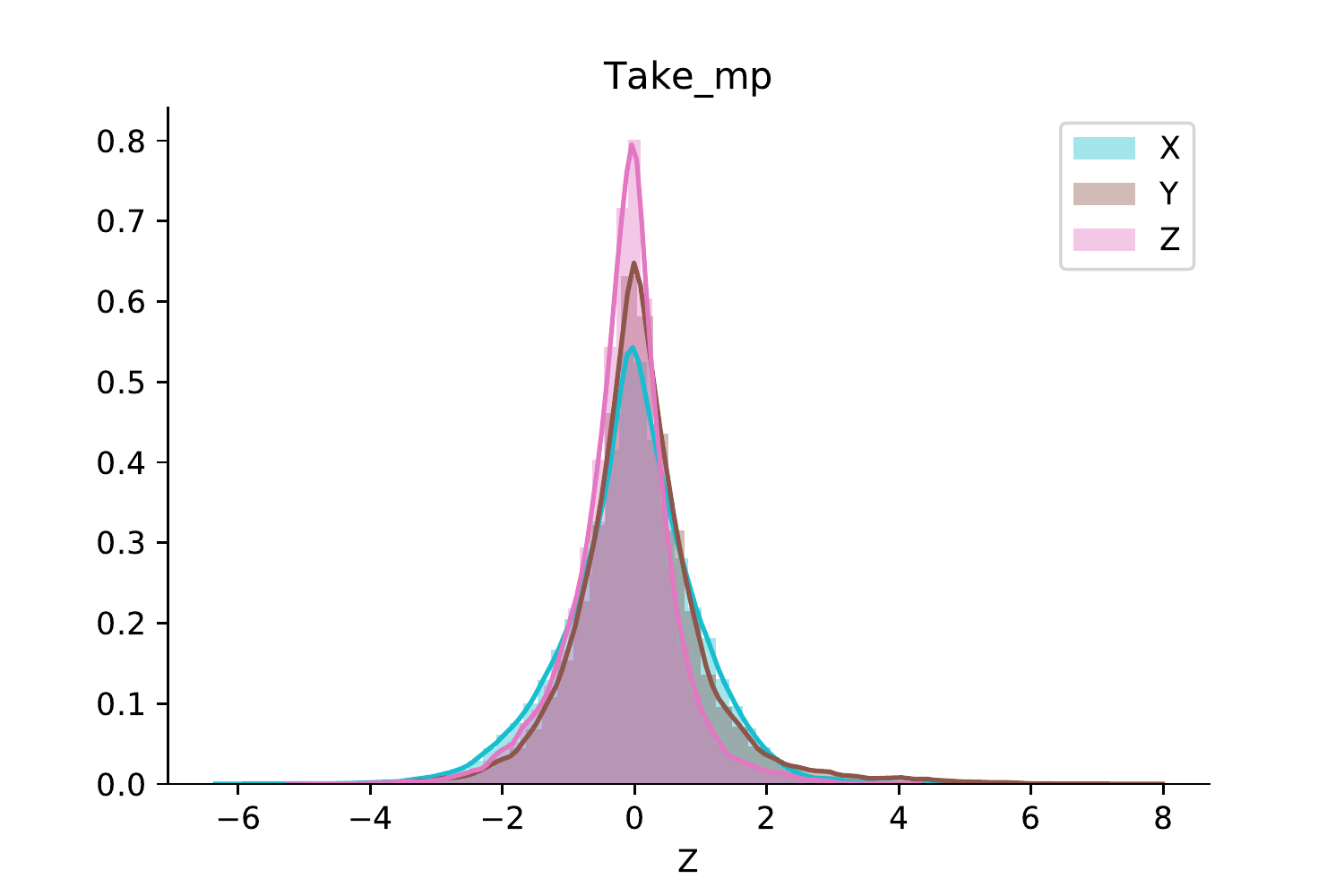}
    \includegraphics[width=0.3\textwidth]{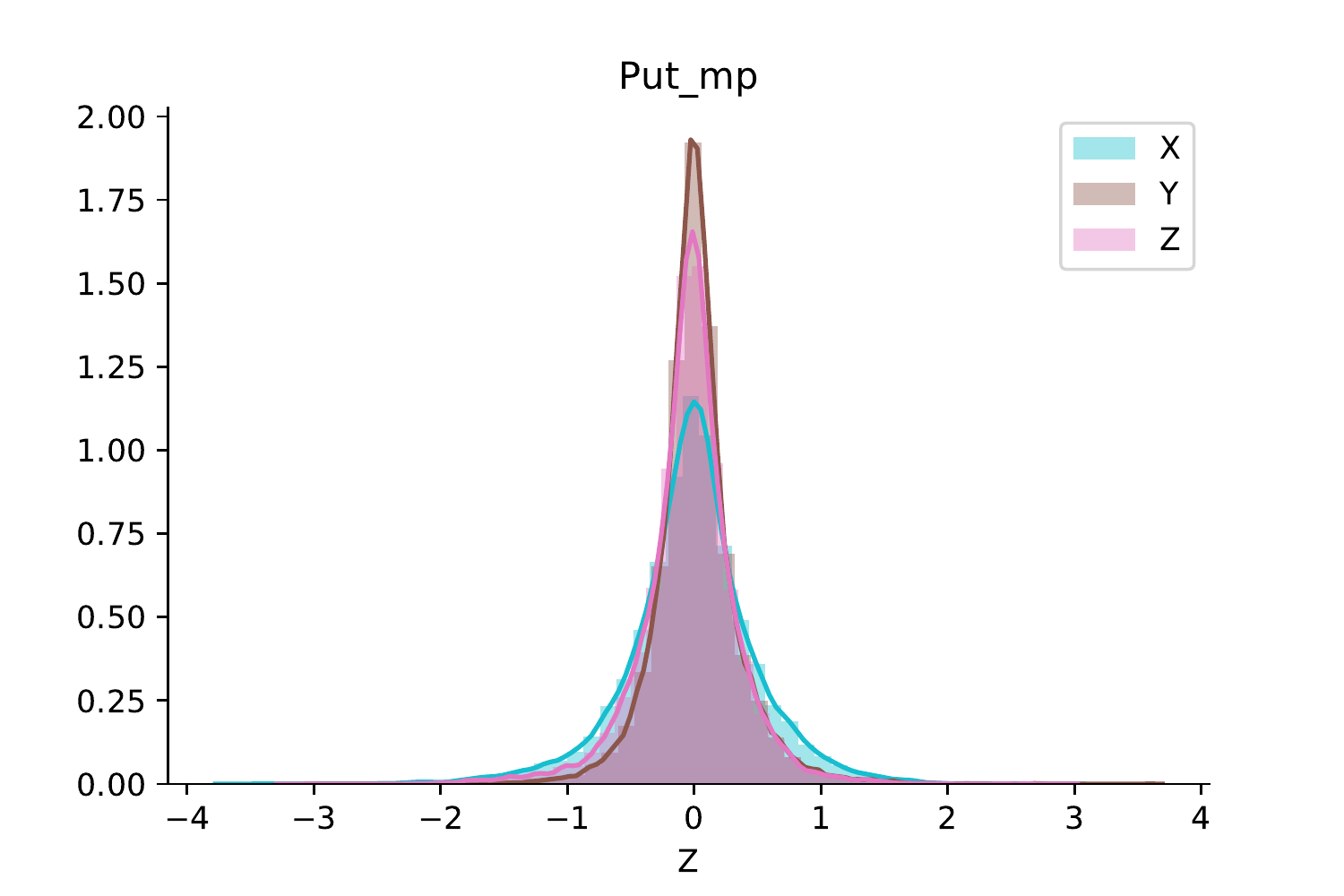}
    \includegraphics[width=0.3\textwidth]{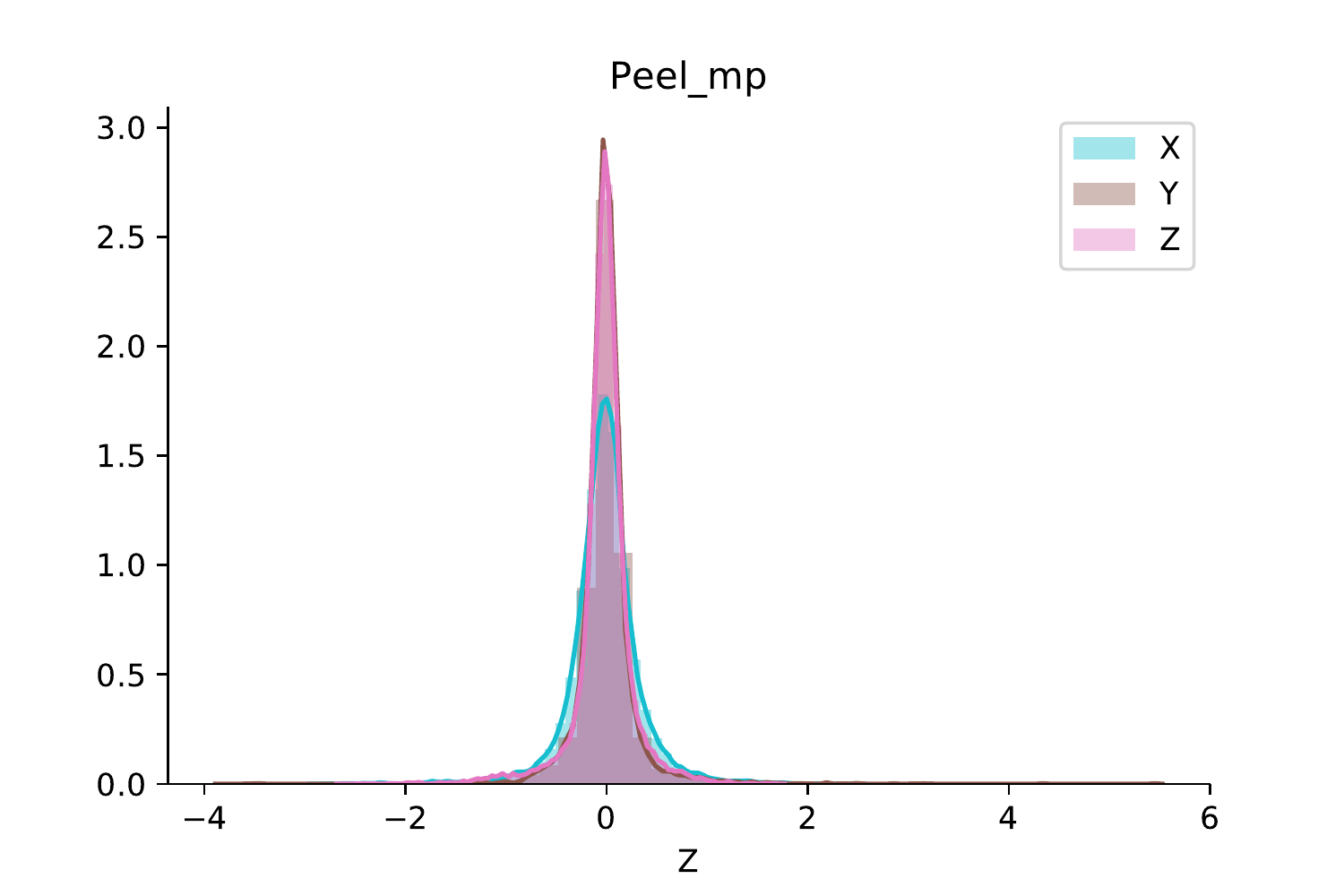}
    \includegraphics[width=0.3\textwidth]{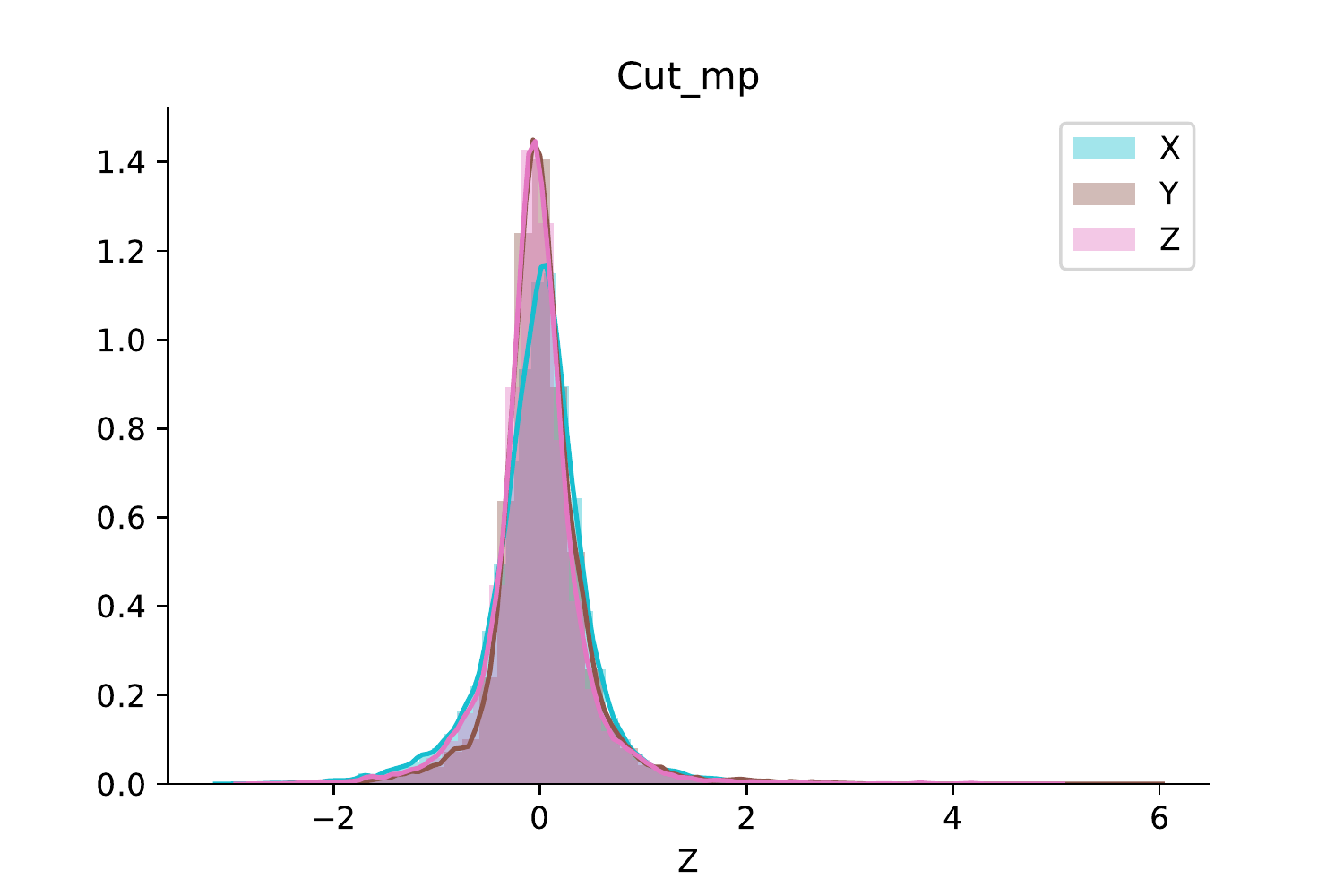}
    \includegraphics[width=0.3\textwidth]{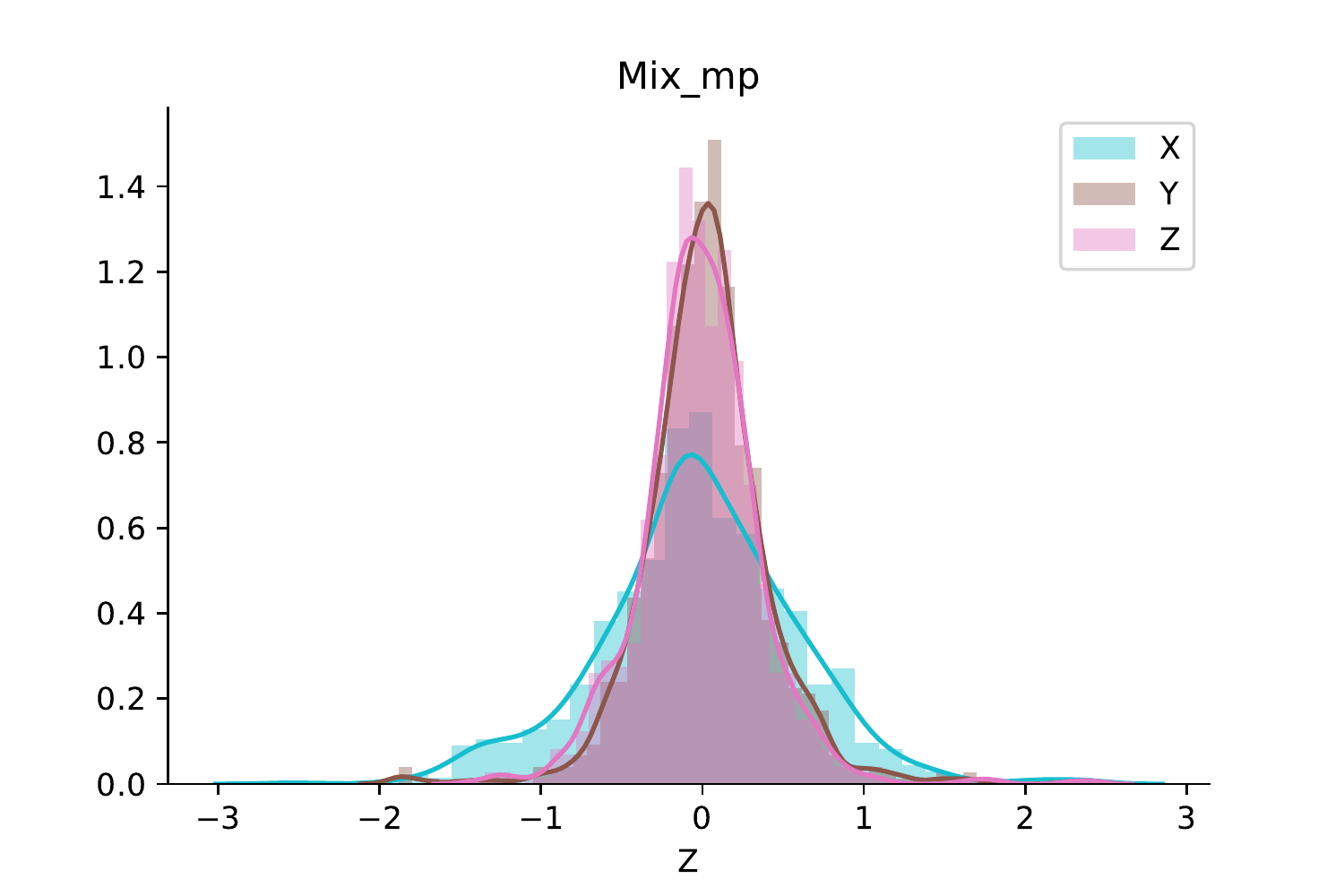}
    \includegraphics[width=0.3\textwidth]{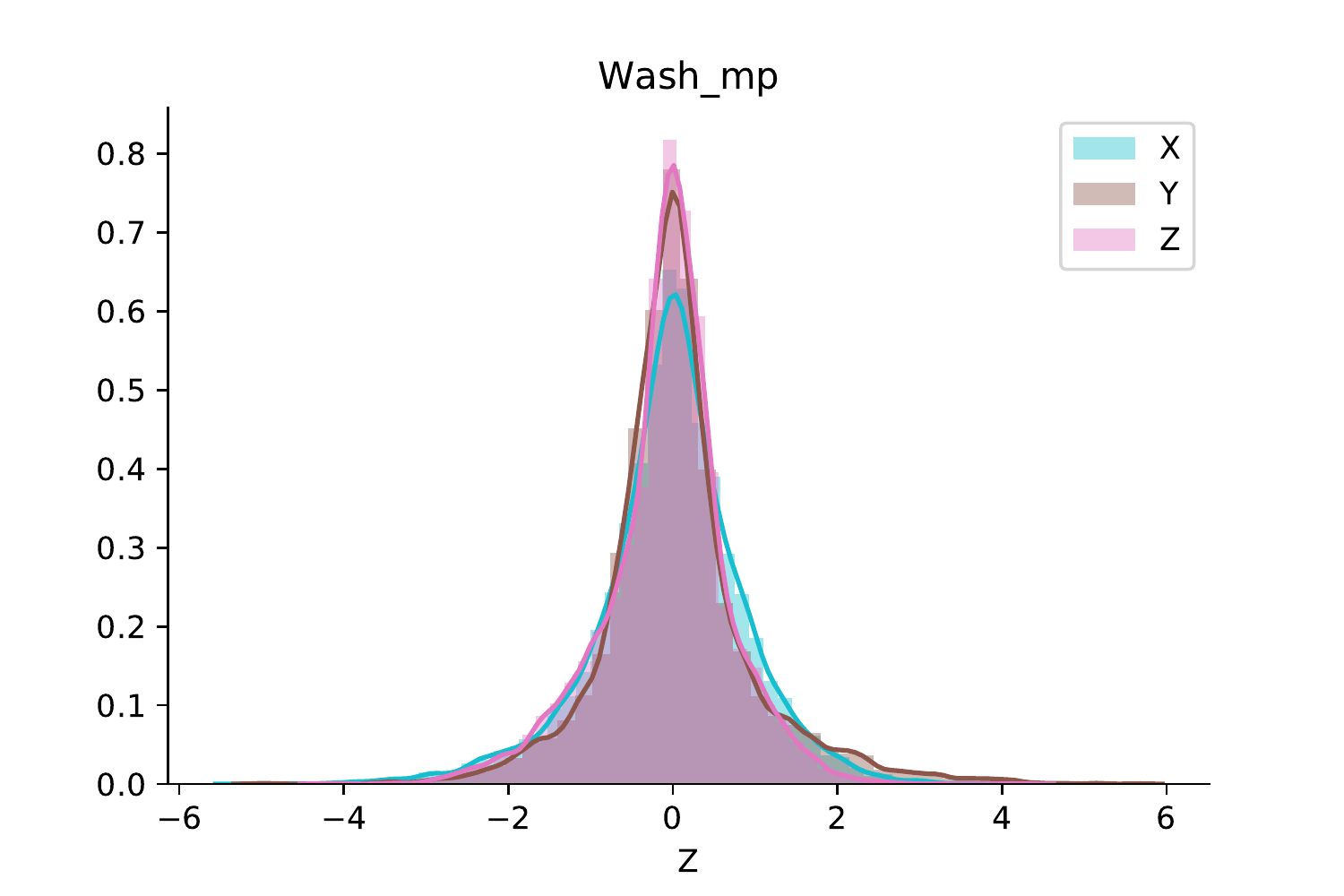}
    \caption{Distribution of the sensor measurements during different micro activities on the fore arm sensor}
    \label{fig:dist_ra}
\end{figure}

\begin{figure}
    \centering
    \includegraphics[width=0.3\textwidth]{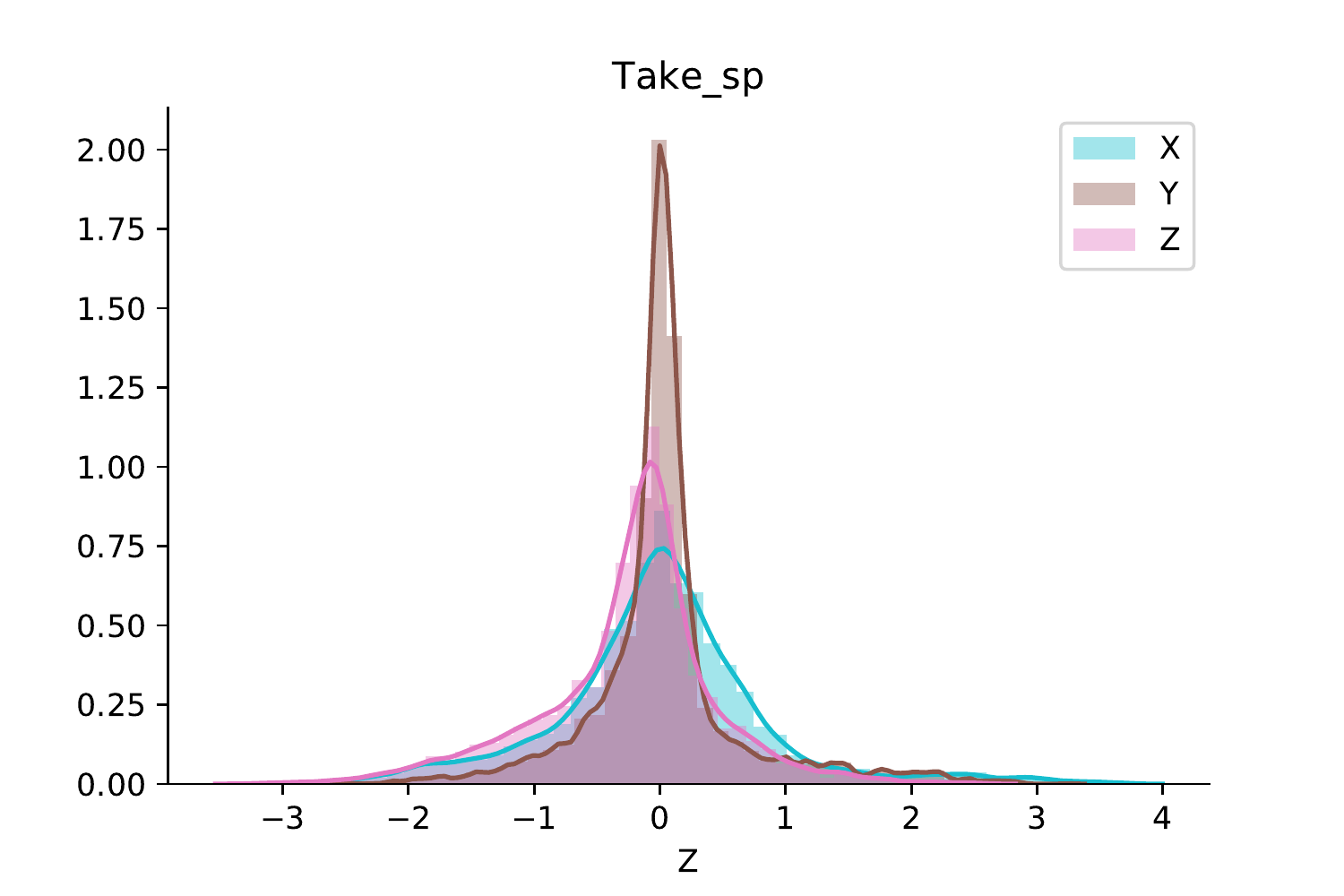}
    \includegraphics[width=0.3\textwidth]{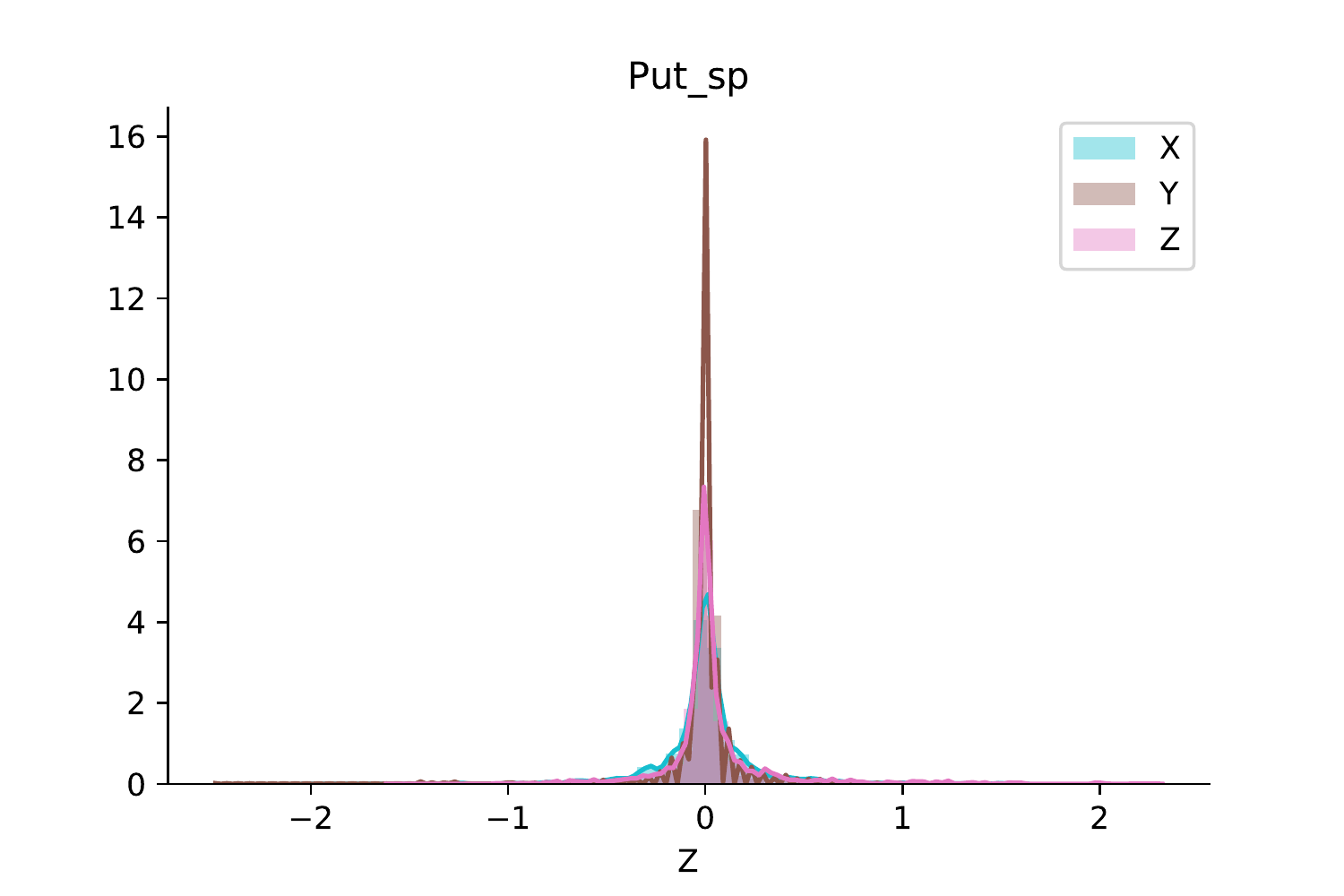}
    \includegraphics[width=0.3\textwidth]{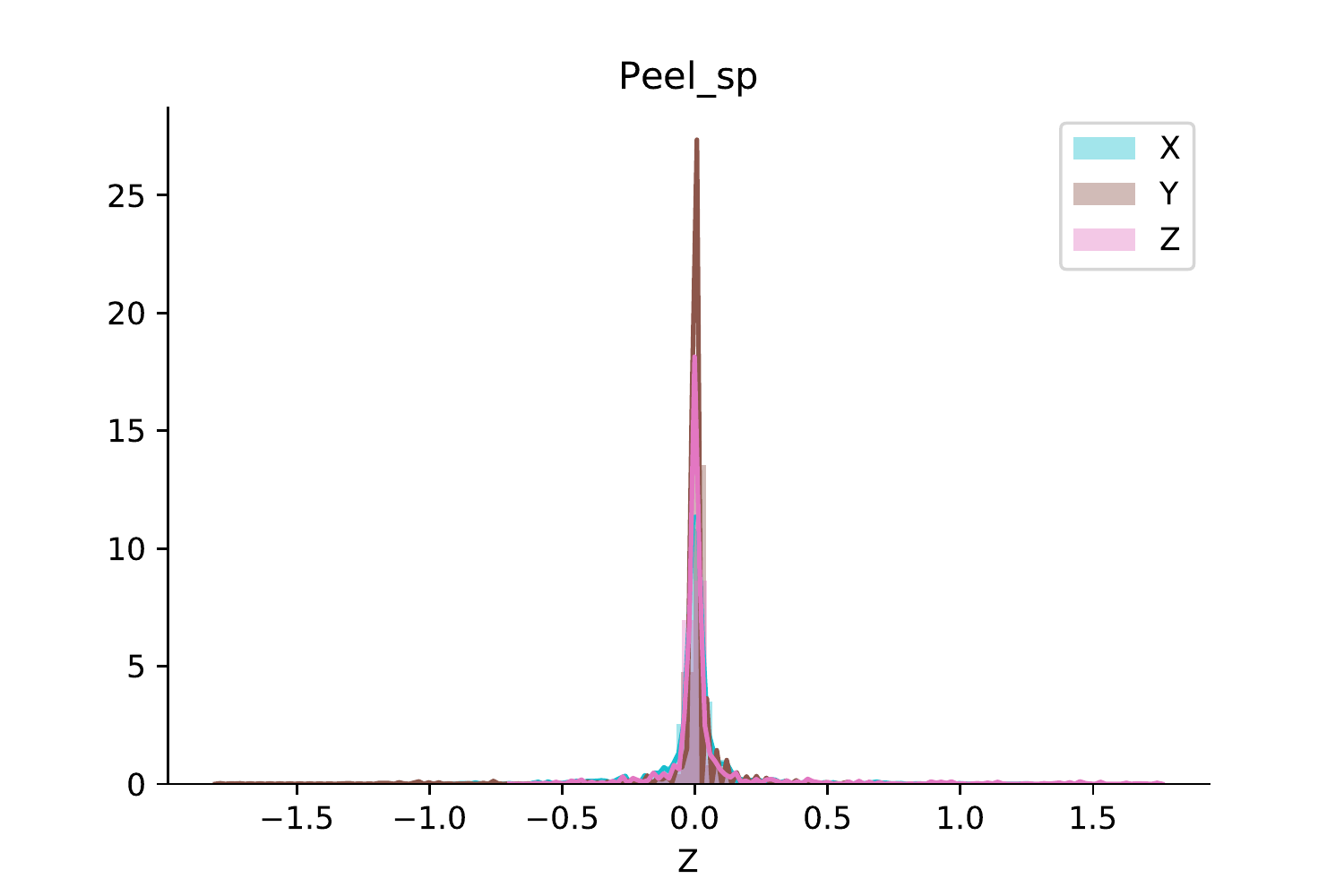}
    \includegraphics[width=0.3\textwidth]{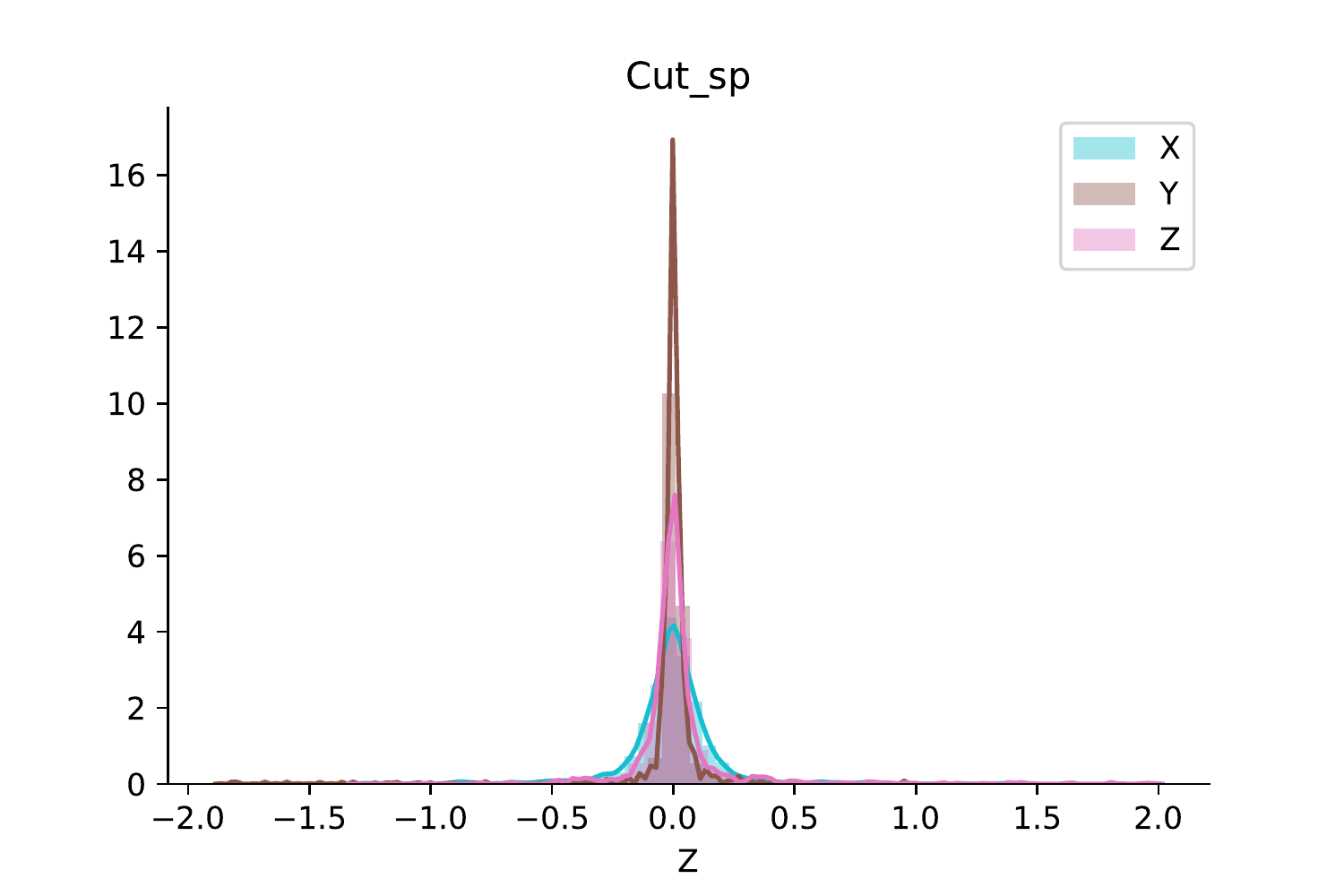}
    \includegraphics[width=0.3\textwidth]{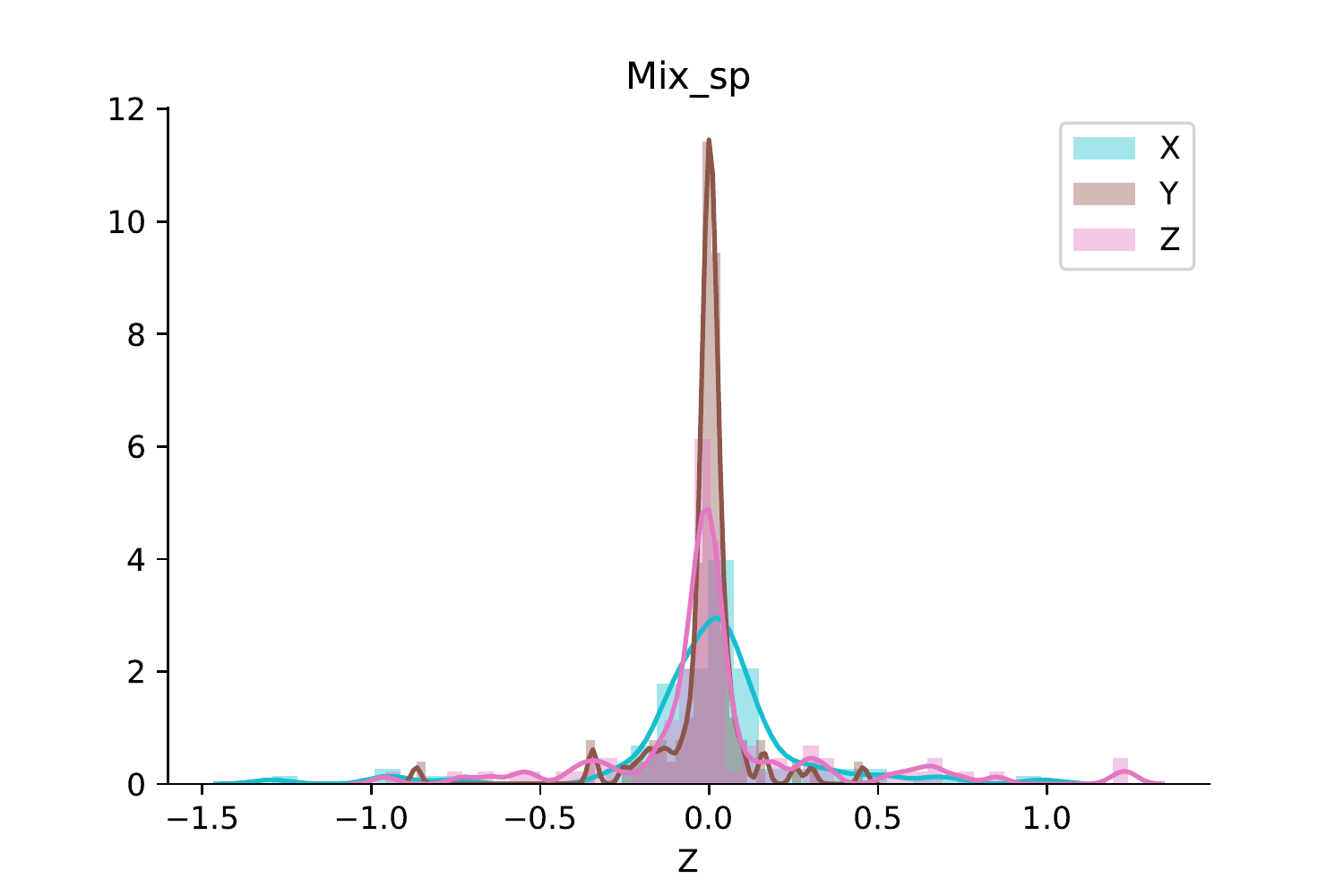}
    \includegraphics[width=0.3\textwidth]{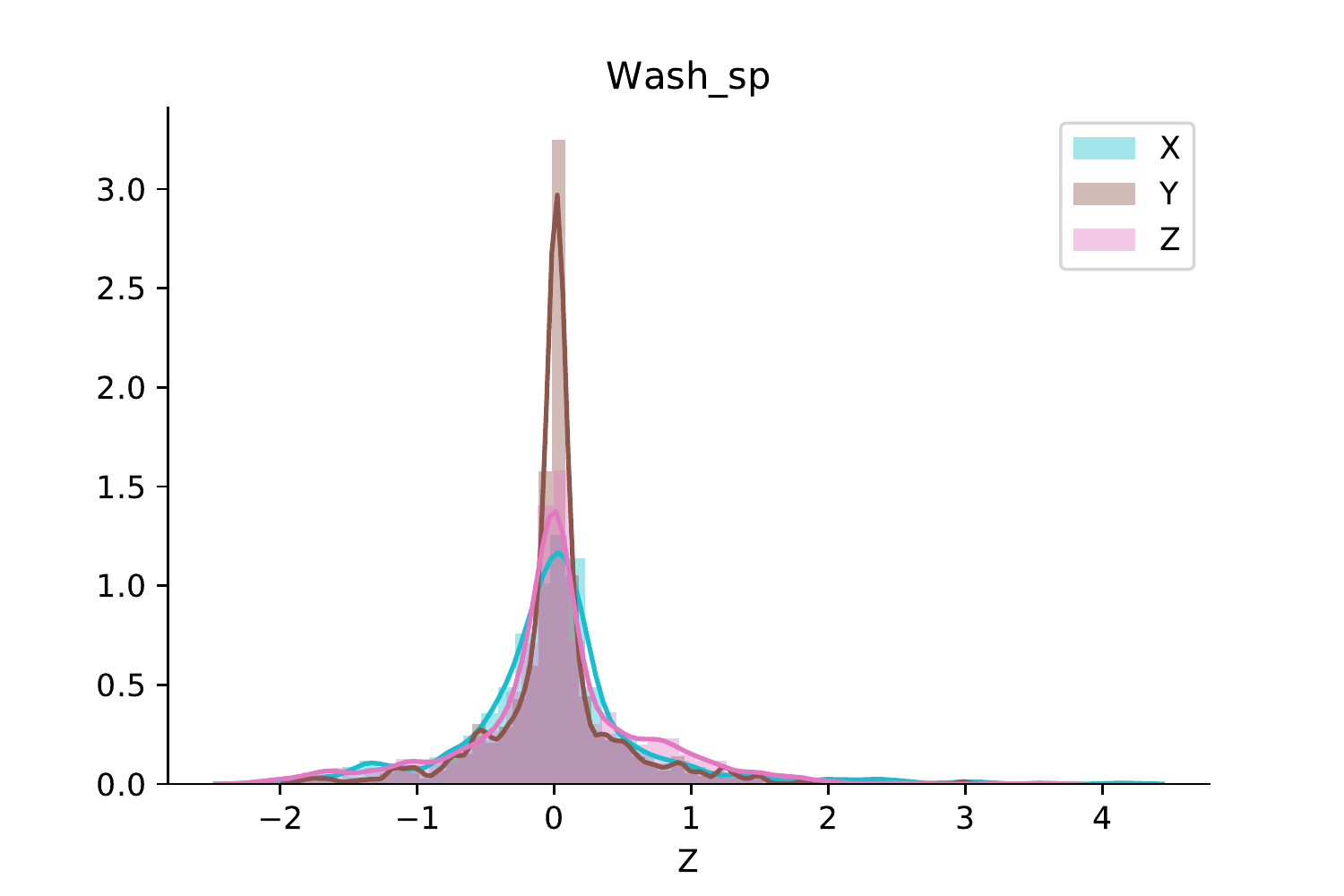}
    \caption{Distribution of the sensor measurements during different micro activities on the hip sensor}
    \label{fig:dist_lh}
\end{figure}

Analyzing the distribution of sensor readings can guide feature selection and help assess whether different activity have different measurement distribution which makes them easier or harder to classify. We show in Figure~\ref{fig:dist_lw} the distribution of left-wrist accelerometer sensor measurements for six different activities and in Figures~\ref{fig:dist_rw}, ~\ref{fig:dist_lh}, ~\ref{fig:dist_ra} those of the right wrist, left hip and right arm sensors respectively. Notice that the distributions of the right wrist sensor have the most significant differences, showing that this sensor might be the most important for classification. In this case, all participants are right-handed which helps explain this result. Although the distributions look similar, their range is different as can be seen from the x-axis limits in the graphs, as is their standard deviation. This suggests that standard deviation, maximum value, minimum value and mode value might be better features than mean value. 

\begin{figure}
    \centering
    \includegraphics[width=0.9\textwidth]{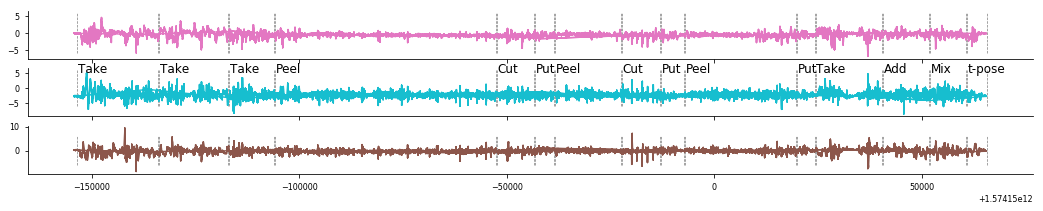}
    \caption{Example of the time series of sensor measurements for the right wrist sensor during fruit salad activity}
    \label{fig:acc_series}
\end{figure}

Notice also that most distribution have a center in zero-value which is explained by the multiple crossings that the signal does (transitioning from positive to negative value) as is shown in Figure~\ref{fig:acc_series}.  

\subsection{Data Quality Analysis}
\label{sec:missing}
Due to the irregular sampling rates caused by Android and some communication problems with bluetooth, we experience some "missing data" problems when the time series is resampled to achieve a regular sampling rate. In this section we analyze the missing data rate when the accelerometer series are resampled at 20Hz. 

We calculate the expected number of non-null samples based on the total time for each micro-activity (based on label times) and obtain the number of non-null samples after resampling~\footnote{we used resampling with a limit of 5 samples, so if more than 5 consecutive null values then it is not interpolated}. 

\begin{figure}
    \centering
    \includegraphics[width=\textwidth]{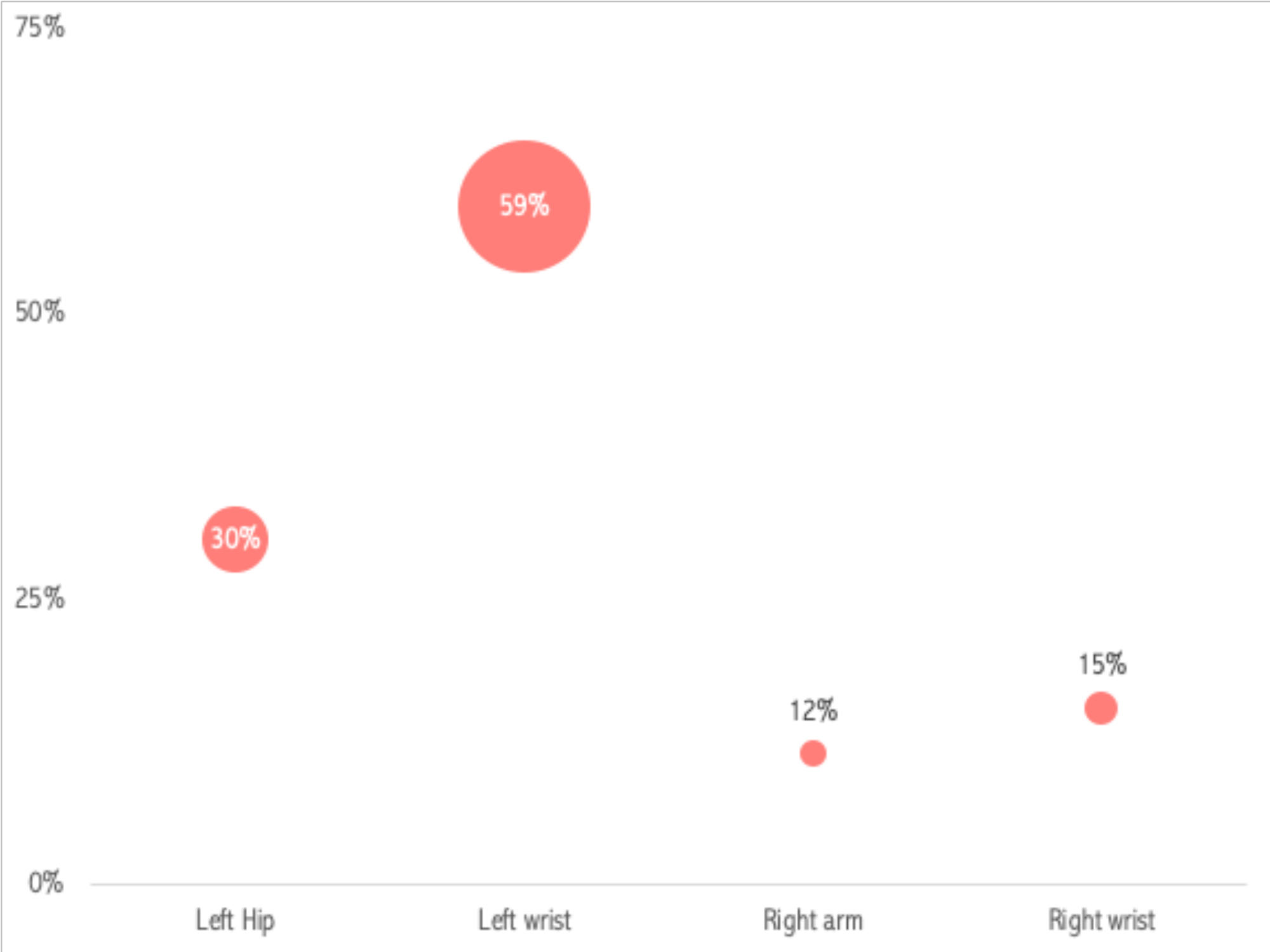}
    \caption{Missing data rate of each accelerometer sensor. Low battery and communication with phone are the main causes of missing data.}
    \label{fig:missing_data}
\end{figure}

Figure~\ref{fig:missing_data} shows the rate of missing samples per sensor. It is evident that the left sensors had a larger missing data rate. There are many possible causes for this. First, communication between the smartwatch and the smartphone may have been interrupted or delayed, causing the data to be missed. Second, when the battery of the smartwatch was low, the sampling rate significantly decreased, causing missing data. Third, operation errors when saving the data  may have caused missing files (each smartphone was operated by a different person). Due to these problems, we don't recommend the use of the left-wrist data but make it available for research in such scenarios. 

Another cause for this missing data is the state of the battery of the smart watch. We notice that as the battery drained, the number of missing data got higher. This is an important consideration for wearable activity recognition, as most reports are based on complete, clean data. In real-life, activity recognition will not achieve the same accuracy if the battery of the device is not high. This is one reason for high- dropout rates of wearable devices.  If we can make models that work well even under low-battery conditions, then user acceptance could potentially increase. 
\section{Activity Recognition Evaluation}
\label{sec:baseline}

An important application is that of recognizing micro-activities. With a correct activity estimation, it is possible to assist a user during cooking or understand if all steps have been followed. In this section, we evaluate activity recognition for the micro-activities in the dataset.  For this, we follow a typical activity recognition pipeline. We evaluate different features, window sizes and algorithms.

\subsection{Activity Recognition Pipeline Description}
As a first step we re-sample all signals to 20Hz. This is to correct for the different sampling rates of all sensors and the sampling rate variability caused by Android API.  

\begin{figure}
    \centering
    \includegraphics[width=0.45\textwidth]{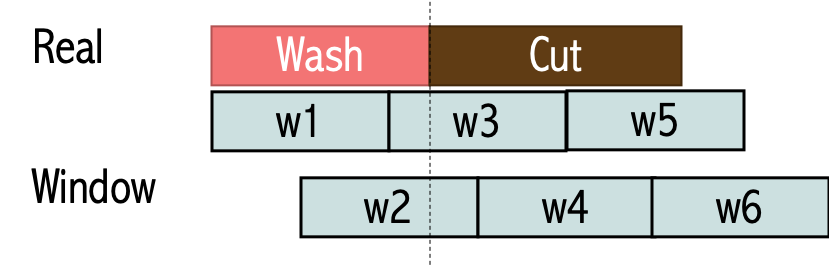}
    \caption{A window may contain traces of different activities. In this example both windows 2 and 3 contain traces of 'Wash' and 'Cut', though at different proportions each. }
    \label{fig:window_labeling}
\end{figure}

 Then we segment the data into sliding windows with 50\% overlap. We use window sizes of 1, 2, 3, 4 and 5 seconds. We chose such sizes considering the average duration of the activities as longer windows would possibly contain traces of a large number of micro-activities~(Figure~\ref{fig:window_labeling}). Even with short windows there might be cases when the situation depicted in Figure~\ref{fig:window_labeling} occurs. In such cases, we chose the label for the window based on the time for each label, such that the label with the longer time is the label for the window. For example, in Figure~\ref{fig:window_labeling}, window 2 is labeled with Wash and window 3, with Cut. 

For each window we extract features using two common feature extraction techniques for activity recognition: 

\textit{Statistical Features}~\cite{Zhang2011, Huynh2005}: summary statistics calculated over the frequency and/or time domains. We use only time-domain features and include mean, standard deviation, maximum, minimum, kurtosis, skew, interquartile range, mean of derivative, and standard deviation of derivative

\textit{Distribution based features}~\cite{Hammerla2013}: Using the Empirical Cumulative Distribution (ECDF) was proposed to capture the characteristics of the distribution of input data. Instead of representing the input as the time sequence values, the cumulative distribution is calculated and the values of the function at equally spaced points are used as features. We use 30 points to represent the ECDF of each axis. 

We evaluated three models: Support Vector Machines with linear and RBF kernel and Random Forests. These models have shown good results in previous literature. 
Although deep learning approaches have also shown high results, we consider that the amount of data in this experiment is not large enough for a deep learning model. 

\subsection{Results}
We evaluated our results using the macro and the micro-average F1-Score. The macro average F1-Score is penalized by the large imbalance in the dataset, however, the micro average reduces the weight of the minority classes. Taking the micro-average results in higher scores largely due to the better recognition of the majority classes. 

\begin{figure}
    \centering
    \includegraphics[width=0.45\textwidth]{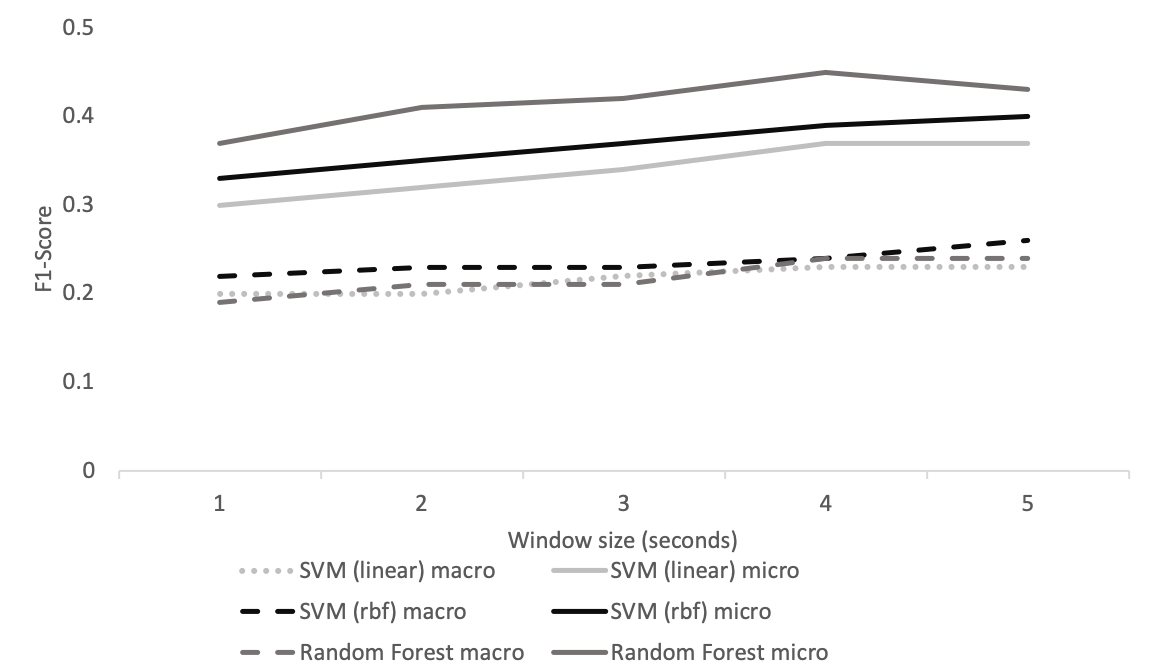}
    \caption{Micro activity classification F1-Score (micro and macro average) for classifiers using statistical features.}
    \label{fig:f1stat}
\end{figure}

\begin{figure}
    \centering
    \includegraphics[width=0.45\textwidth]{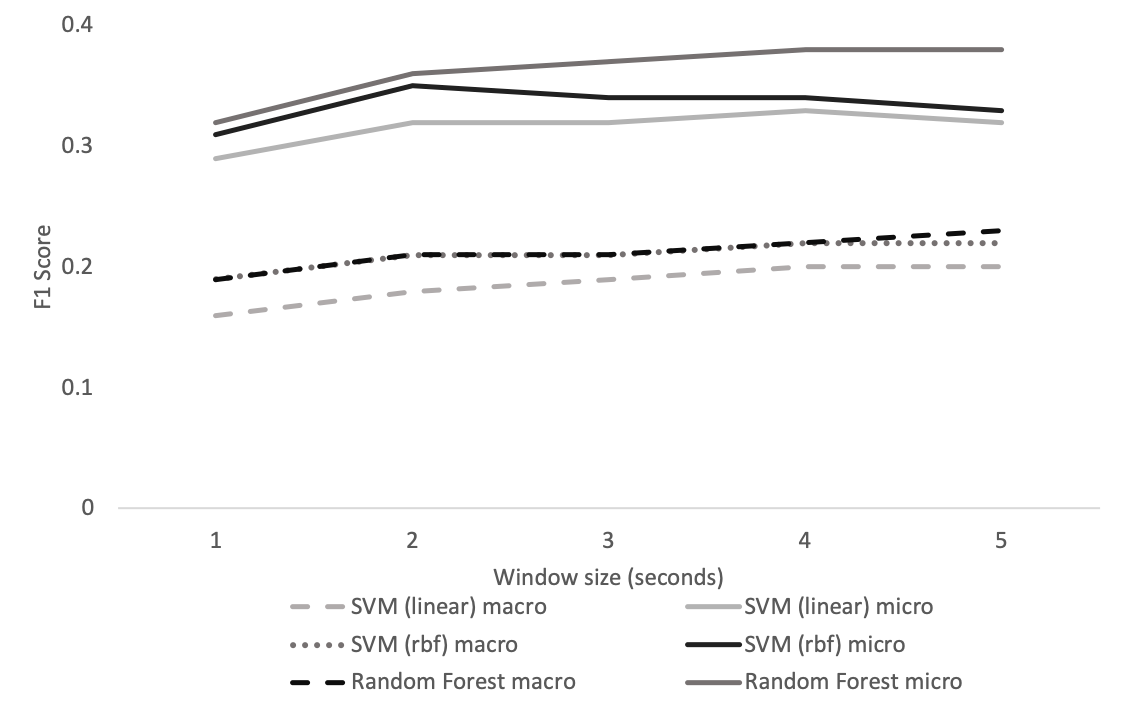}
    \caption{Micro activity classification F1-Score (micro and macro average) for classifiers using ECDF features.}
    \label{fig:f1ecdf}
\end{figure}

\begin{figure}
    \centering
    \includegraphics[width=0.9\columnwidth]{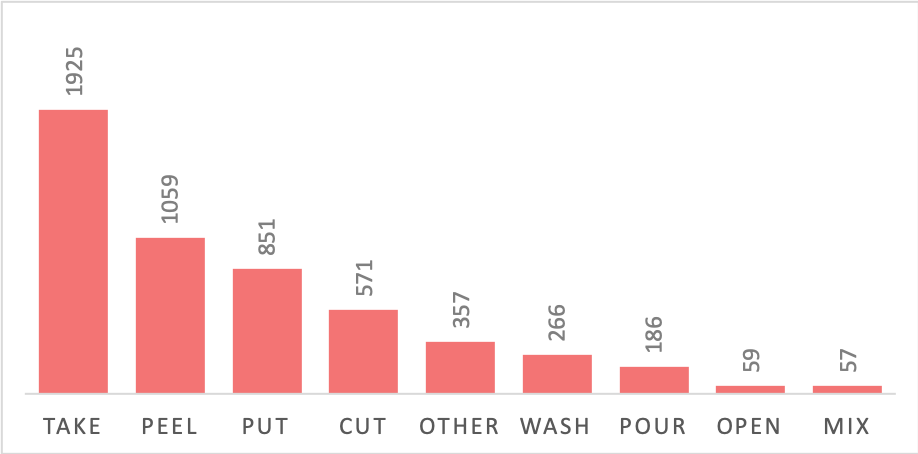}
    \caption{Number of windows per micro-activity when window length is 4 seconds.}
    \label{fig:windows-4seconds}
\end{figure}

We evaluated three classifiers with two different feature sets each. The results for all classifiers using the statistical features and the ECDF features are shown in Figures~\ref{fig:f1stat} and~\ref{fig:f1ecdf}, respectively. The number of windows for each activity when using 4 seconds is shown in Figure~\ref{fig:windows-4seconds}

We observe that, for all classifiers, the statistical features perform better than the Empirical Cumulative Distribution Features. Due to the imbalance, the micro-average F1-Score is higher than the macro-average score. As reference, a classifier that classifies all windows in the majority class (Take) would achieve a macro F-Score of ~6\% and a micro F-Score of 36\%~\footnote{when the window size is 5 seconds.}. Trained models show a better macro F-Score but little improvement for the micro score.  

We also observe that the peak performance is usually at windows of 4 seconds. Nonetheless, considering the average duration of the micro-activities~(table~\ref{tab:avg_duration}), when 4 seconds windows are used the shortest activities might be hard to distinguish because they have a high probability of belonging in window with other activity. 


\section{Conclusions}
\label{sec:learned_rec}

In this paper we have introduced a new dataset for activity recognition based on body-movement analysis that combines visual and inertial sensor data. The dataset has been labeled at two different granularity labels: macro activities, that represent different recipes; and micro activities, representing the steps followed for preparing the recipes. The combination of these two levels can provide researchers with data to study algorithms that combine recognition of actions, shared by many activities, and activities, composed of similar actions in different orders or frequencies. 

This dataset also combines visual and inertial sensors for measuring body movement. For both sensors, we used two different types of sensors. As for the visual sensors, we have used motion capture and open pose. As for the inertial measurement sensors, we used smartphones and smartwatches. This combination also allows researchers to experiment and measure model performance degradation when data quality is lower. In addition, as there are multiple body positions being tracked by each type of sensor, performance under different combinations of sensors can also be studied. 

We have described the protocol for data collection. Our interest was to collect realistic data. Therefore we used commercial smartphones and smartwatches instead of specific-purpose sensors which can be more accurate and have a constant sampling rate.  We observed a high-missing data rate for the left smartwatch, despite it being the same model of the right smartwatch. The missing data rate is an important consideration when designing applications for real-life use. 

The 'Cooking Dataset' introduced in this paper proposes several challenges for the activity recognition community. Not only are there two levels of granularity in the labels, but also there is a high imbalance in the micro-activity level. This imbalance comes from the different duration and frequency of the activities. Understanding how this play a role in the metrics and in final applications is an interesting research direction that we would like to study.

\section*{Acknowledgment}
The authors would like to thank the participants of the experiment for their time and collaboration during data collection.

\bibliographystyle{plain}
\bibliography{references}

\section*{Appendix}
\label{sec:appendix}

\end{document}